\documentclass[a4paper]{article}

\newcommand{\ww}{.75} 

\usepackage{amsmath,amssymb}

\usepackage[margin=1.25in]{geometry}
\usepackage{microtype}
\usepackage[colorlinks = true,
linkcolor = blue,
urlcolor  = blue,
citecolor = blue,
anchorcolor = blue]{hyperref}
\usepackage{graphicx}
\usepackage{tabularx}

\usepackage[caption=false]{subfig}
\usepackage{floatrow}
\usepackage{siunitx}
\usepackage[table]{xcolor}

\usepackage{ulem}
\usepackage{dcolumn}

\usepackage{todonotes}
\usepackage{enumitem}

\usepackage{multirow}
\usepackage{bm}
\usepackage[most]{tcolorbox}
\usepackage{authblk}

\DeclareSIUnit{\molar}{M}

\newcommand{\yp}[1]{{\color{black}{#1}}}

\newcommand{\ya}[1]{{\color{black}{#1}}}
\newcommand{\yc}[1]{{\color{black}{#1}}}

\newcommand{\ve}{\varepsilon}
\newcommand{\pa}{\partial}

\newcommand{\p}{\mathbf{p}}

\newcommand{\x}{\bm{x}}
\newcommand{\ub}{\bm{u}}
\newcommand{\rss}{\text{RSS}\xspace}

\newcommand\no{N\textsuperscript{\underline{\scriptsize o}}\xspace}

\newcommand{\mcomment}[1]{}

\usepackage{xspace}
\newcommand{\tar}{trap-and-release\xspace}

\newcommand{\bko}{\cellcolor{gray!20}}
\newcommand{\bkp}{\cellcolor{violet!20}}
\newcommand{\bkg}{\cellcolor{brown!20}}

\newcommand{\cko}{\cellcolor{gray!20}\checkmark}
\newcommand{\ckp}{\cellcolor{violet!20}\checkmark}
\newcommand{\ckb}{\cellcolor{brown!20}\checkmark}

\usepackage{pifont}
\newcommand{\xx}{\ding{55}\xspace}

\usepackage{tikz}
\usetikzlibrary{fit,shapes.misc}

\date{}

\begin{document}
	
	\title{Models of Vimentin Organization Under Actin-Driven Transport}
	
	\author[1]{Youngmin Park\footnote{Corresponding author \href{mailto:park.y@ufl.edu}{park.y@ufl.edu}. Current address: University of Florida Department of Mathematics 1400 Stadium Rd Gainesville, FL 32611 USA}}
	\author[2]{C\'{e}cile Leduc}
	\author[3]{Sandrine Etienne-Manneville}
	\author[1]{St\'{e}phanie Portet}
	
	\affil[1]{\small University of Manitoba Department of Mathematics
		420 Machray Hall, 186 Dysart Road University of Manitoba, Winnipeg, MB R3T 2N2 Canada}
	\affil[2]{\small Université Paris Cité, CNRS, Institut Jacques Monod, F-75013 Paris, France}
	\affil[3]{\small Cell Polarity, Migration and Cancer Unit, Institut Pasteur, UMR3691 CNRS. Equipe Labellis\'ee Ligue Contre le Cancer, I-75015, Paris, France}
	

	
	
	
	
	\maketitle
	
	\begin{abstract}
		Intermediate filaments form \yp{an essential structural network\yp{,} spread throughout the cytoplasm and play a key role in cell mechanics, intracellular organization and molecular signaling. The maintenance of the network and its adaptation to the cell's dynamic behavior relies on several mechanisms implicating cytoskeletal crosstalk which are not fully understood. Mathematical modeling allows us to compare several biologically realistic scenarios to help us interpret experimental data.} In this study, we observe and model the dynamics of the \yp{vimentin intermediate filaments in single glial cells} seeded on circular micropatterns \yp{following microtubule disruption by nocodazole treatment. In these conditions, the vimentin filaments move towards the cell center and accumulate before eventually reaching a steady-state. In absence of microtubule-driven transport, the motion of the vimentin network is primarily driven by actin-related mechanisms.} \yp{To model these experimental findings, we hypothesize that vimentin} may exist in two states, mobile and immobile, and switch between the states at unknown \yp{(either constant or non-constant)} rates. Mobile vimentin are assumed to advect with \yp{either} constant or non-constant velocity. We introduce several biologically realistic {scenarios} using this set of assumptions. For each {scenario}, we use differential evolution to find the best parameter sets resulting in a solution that most closely matches the experimental data, then \yp{the assumptions are evaluated} using the Akaike Information Criterion. \yp{This modeling approach allows us to  conclude that our experimental data are best explained by a spatially dependent trapping of intermediate filaments \yp{or a spatially dependent speed of actin-dependent transport.}}
	\end{abstract}
	
	\section{Introduction}
	
	Intermediate filaments are key components of the cytoskeleton and are involved in  fundamental cell functions including stress response, cell growth, proliferation, migration and death \cite{SEM_review2018,eldirany2021recent,nunes2021molecular}.
	\yp{The organization of the cytoplasmic network formed by} intermediate filaments \yp{endows the cell with robust mechanical properties \cite{ramms2013keratins,latorre2018active,van2021intermediate}, and is critical for all intermediate filament functions.} Disruption of intermediate filament organization in cells is observed in numerous diseases related to mutations of intermediate filament proteins \cite{fuchs1998structural,omary2009if,coch2016intermediate}. This relationship between {organization} and function warrants a careful study of the primary drivers behind {the dynamic spatial distribution of} intermediate filaments.
	
	Intermediate \yp{filament proteins} organize to form \yp{a dynamic filamentous network} through three interdependent processes: one, assembly and disassembly of soluble intermediate filament proteins \cite{omary2006heads,hookway2015microtubule}, two, active transport of filaments via molecular motors walking on microtubules or actin {fibers} \cite{gyoeva1991coalignment,prahlad1998rapid,helfand2003rapid,leduc2017regulation}, and three, a \yp{continuous retrograde flow towards the cell center that affects intermediate filament organization,} resulting from \yp{centripetal movement of actin filaments, in part powered by acto-myosin contractility} 
	\cite{hollenbeck1989intermediate,kolsch2009actin,dupin2011cytoplasmic,jiu2015bidirectional,leduc2017regulation}. 
	\yp{In  \cite{jiu2015bidirectional}, plectin-mediated crosslinks between actin and vimentin intermediate filaments are shown to affect the organization of both cytoskeletal systems.} These processes interact in a constant state of flux, helping to maintain homeostasis of the intermediate filament network in interphasic cells under no cytopathogenic conditions. However, there is little known about how each process contributes and interacts towards intermediate filament network spatial distribution and organization.
	
	Mathematical modeling studies often explore specific elements of the above processes, including mechanisms for intermediate filament \textit{in vitro} assembly and disassembly \cite{kirmse2007quantitative,portet2009vimentin,portet2013dynamics,martin2015model,mucke2016vitro,mucke2022general,SCHWEEN20223850,PhysRevX.13.011014}, intermediate filament \yp{\textit{in vivo}} network formation and organization \cite{portet2003organization,beil2009simulating,sun2015mathematical,sun2017mathematical,portet2015keratin,gouveia2021}, and intermediate filament transport along microtubules driven by motor proteins such as kinesin and dynein \cite{craciun2005,brown2005,Kuznetsov2013,li2014,lee2015,dallon2019stochastic,portet2019deterministic,portet2022noise,BrowneN2022,Dallon_PLOSCOMPBIO2022}. Modeling studies that consider interactions between actin and intermediate filament networks are limited to the properties of the resulting network, such as \cite{lopez2019theory}, where the authors characterize the robustness of the intermediate filament network with and without actin, and \cite{kim2012mathematical}, where the authors find that the interaction between actin and intermediate filaments control the extent of keratinocyte cell spreading.
	
	In contrast to existing studies, our long-term goal is to understand how the processes interact to form and maintain intermediate filament networks. Some work has been done in this direction, e.g., \cite{portet2015keratin}, where the interplay between a net inward transport and assembly/disassembly processes is considered and the net transport is found to be the dominating process. In the present study, we simplify the problem by eliminating one of the three processes by applying nocodazole \yp{to depolymerize microtubules,} disrupting microtubule-dependent \yp{transport} (kinesins and dyneins). What remains is actin-\yp{mediated transport} \ya{of intermediate filaments. The resulting} experimental data consist of intermediate filament spatial distributions over time in primary astrocytes (major glial cells of the central nervous system). In particular, the distribution data are of fluorescent vimentin, which are an intermediate filament protein expressed in mesenchymal origin cells. Modeling this data allows us to infer the underlying biological mechanisms \ya{of intermediate filament organization}.
	
	
	\yp{All code and data used to generate figures are publicly available on GitHub at
		\url{https://github.com/youngmp/retrograde_flow_models}}
	
	
	\section{Methods}
	
	\subsection{Experimental Protocol}
	
	\paragraph{Cell culture}
	Primary rat astrocytes were prepared as previously described in \cite{etienne2006vitro}, according to the guidelines approved by the French Ministry of Agriculture, following European standards. Cells were grown to confluence in Dulbecco's Modified Eagle Medium medium with \SI{1}{g/L} glucose and supplemented with 10$\%$ FBS (Invitrogen, Carlsbad, CA), 1$\%$ penicillin–streptomycin (Gibco, ThermoFisher scientific) and 1$\%$ Amphotericin B (Gibco, ThermoFisher scientific).
	
	\paragraph{Micropatterns and drug treatment}
	The micropatterning technique is used to impose reproducible cell shape and decipher cell morphogenesis and functions \cite{thery2010micropatterning}. We have previously used astrocytes plated on micropatterns to study cell polarization \cite{leduc2017regulation}. Briefly, primary rat astrocytes are plated onto glass-bottom tissue culture dishes coated with fibronectin after deep UV micropatterning of the surrounding polyethylene glycol (PEG). We used \SI{60}{\um}-diameter disks where only single cells were allowed to spread. Cells cannot adhere outside of the micropattern. Nocodazole from a stock \SI{10}{\milli \molar} in DMSO (Sigma Aldrich) was added to the cells at a final concentration of \yp{\SI{10}{\micro \molar}}.
	
	\paragraph{Immunofluorescence}
	Cells were fixed in cold methanol for 5 minutes and blocked with 3$\%$ BSA in PBS for \SI{1}{h}. Cells were then incubated for 1 hour with vimentin primary antibodies (Santa Cruz Biotechnology \no sc-7557R) diluted 50 times in PBS, washed three times in PBS and then incubated another hour with secondary antibodies (Jackson Immuno Research, Alexa Fluor 488) diluted 500 times in PBS. Finally, coverslips were washed and mounted in Prolong Gold with DAPI (Thermo fisher). Epifluorescence images were obtained on a microscope (model DM6000, Leica, Solms, Germany) equipped with 40x, NA 1.25 and a 63x, NA 1.4 objective lenses and were recorded on a CCD camera using Leica software. 
	
\begin{figure}
		\includegraphics[width=\ww\textwidth]{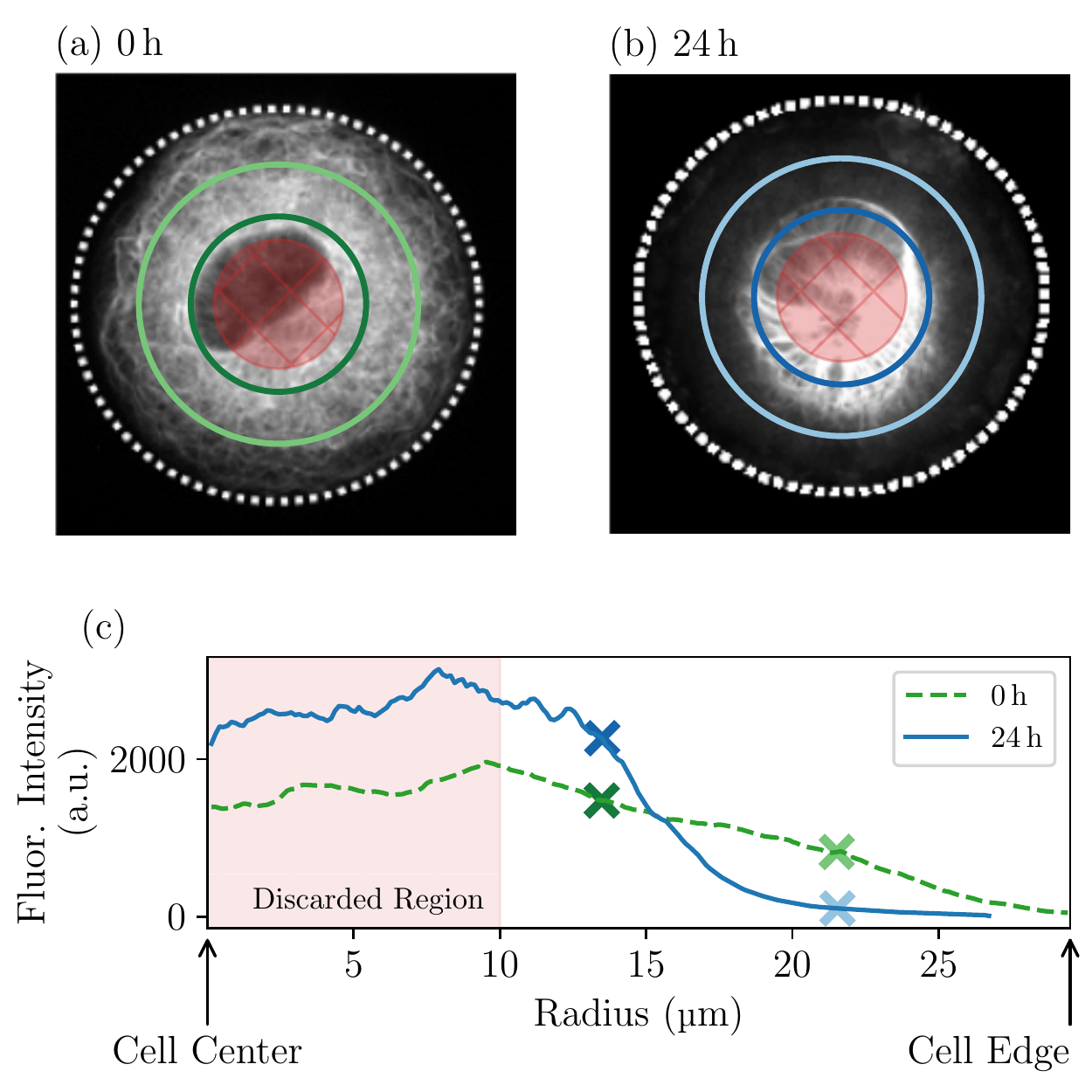}
		\caption{Fluorescent vimentin in cells seeded on circular micropatterns. Micropattern areas are shown by the white dotted curves in panels (a), (b). Spatial distribution of fluorescent vimentin in representative cells ({white material within the micropattern boundary} (a), (b)) exhibit an approximate circular symmetry. Fluorescence intensity corresponds directly to vimentin quantity. Red circular regions denote the approximate location of cell nuclei across all cells. (a): Fluorescence intensity of \yp{a} control cell (i.e., cells not subjected to nocodazole), which we treat as an initial condition, hence the label \SI{0}{h}. (b): Fluorescent vimentin of a cell after being exposed to nocodazole for \SI{24}{h}, which we treat as a steady-state profile. In (a) (resp. (b)), vimentin intensity is averaged along concentric circles, with two representative circles shown in green (resp. blue) and their corresponding average intensity values marked by green (resp. blue) x's in panel (c). (c): Average fluorescence intensity curves in arbitrary units (a.u.) as a function of radius from the cell center up to the cell edge. The green dashed curve corresponds to the control cell (a) and the blue curve corresponds to the cell subjected to nocodazole for 24 hours (b). Data in the red region -- corresponding to the red circles in panels (a) and (b) -- are discarded.}\label{fig:experiment}
	\end{figure}

	\paragraph{Radial profile intensity}
	Radial profile intensity was plotted for every cell using the Radial Profile Plot plugin from Fiji {(FIG. \ref{fig:experiment}(c))}. The plugin plots the average intensity around concentric circles as a function of distance from a point in the center of the micropattern. Two representative circles are shown in FIG. \ref{fig:experiment} for a cell before the addition of nocodazole (a, green) and for another cell \SI{24}{h} after the addition of nocodazole (b, blue). The corresponding average intensity values along these circles are marked by x's in panel (c) in corresponding colors.
	
	\subsection{Data}
	
\begin{figure}
		\includegraphics[width=\ww\textwidth]{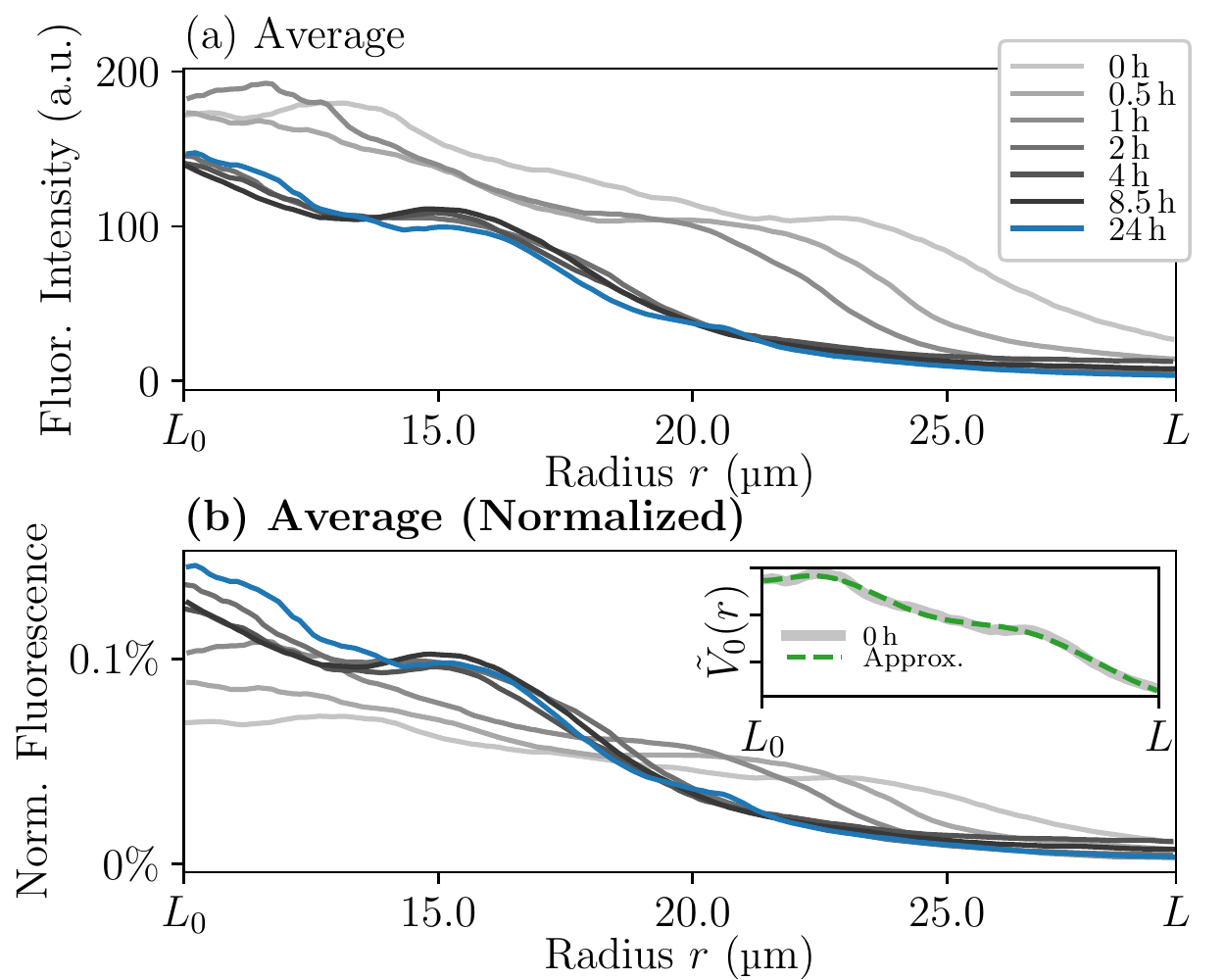}
		\caption{Vimentin fluorescence data over space and time. In all panels, darker shades correspond to later experimental times. The final data at 24 hours is denoted by a blue curve. The $r-$axis of each plot is the domain and corresponds to the radial distance from the cell center. (a): Average vimentin fluorescence profiles. The average data was collected from 4 cells at \SI{0}{h}, 8 cells at \SI{0.5}{h}, 11 cells at \SI{1}{h}, 5 cells at \SI{2}{h}, 11 cells at \SI{4}{h}, 13 cells at \SI{8.5}{h}, and 5 cells at \SI{24}{h}. (b): Normalized average vimentin fluorescence profiles \eqref{eq:normalize}. We use the \SI{0}{h} curve as the initial condition for simulations (written $\tilde V_\text{0}(r)$, where  $r\in[L_0,L]$, for $L_0=\SI{10}{\um}$ and $L=\SI{30}{\um}$). The bold label for panel (b) is to emphasize that we use the normalized, average vimentin data for model-fitting. {(b), inset: Least-squares approximation of data using a sum of 10 Gaussians. Control data (gray) is shown with its approximation (dashed green) superimposed. Free parameters for each Gaussian are the width, amplitude, and shift. We use the approximated curve (dashed green) to initialize model simulations.}}\label{fig:data}
	\end{figure}

	Our fluorescence intensity data is circularly symmetric after computing the radial profile intensity, thus {information} over the radial coordinate {is sufficient to represent data} (FIG. \ref{fig:experiment}(c)). We then average the intensity data over multiple cells that have been exposed to nocodazole for different amounts of time before being fixed (FIG. \ref{fig:data}(a)). The average data come from 4 cells at \SI{0}{h}, 8 cells at \SI{0.5}{h}, 11 cells at \SI{1}{h}, 5 cells at \SI{2}{h}, 11 cells at \SI{4}{h}, 13 cells at \SI{8.5}{h}, and 5 cells at \SI{24}{h}. The red circular regions in FIG. \ref{fig:experiment}(a) \yp{and} (b) denote the approximate region of the cell nucleus, which varies from cell to cell, and is therefore excluded in the model calibration. From this point forward, we do not consider the discarded region for radii between \SIrange{0}{10}{\um} {(FIG. \ref{fig:experiment}(c))}.
	
	Before deriving the model, we formalize notation. Let $L_0=\SI{10}{\um}$ denote the approximate boundary of the nuclear envelope for all cells, and let $L=\SI{30}{\um}$ denote the cell edge. The index set of nocodazole exposure {duration} is given by
	\begin{equation*}
		\mathcal{V}_t := \{0,0.5,1,2,4,8.5,24\} \quad \text{(units in hours)}.
	\end{equation*}
	The index set of radial coordinates is denoted by ${\mathcal{V}}_r$, where the elements are simply the horizontal coordinates of the data curves {(FIG. \ref{fig:data})}.
	
	Let the function $\tilde V_{t}(r)$ represent fluorescence intensity data for hour $t$ at radial coordinate $r$. For each time point, we normalize the data by requiring that
	\begin{equation}\label{eq:normalize}
		2\pi\int_{L_0}^L \tilde V_{t}(r) r \,\mathrm{d}r = 1,
	\end{equation}
	which is simply a normalization using the total area under the averaged data on the micropattern (excluding the \SI{10}{\um}-radius circle) in polar coordinates. We abuse notation and also let $\tilde V_{t}(r)$ denote the normalized fluorescence data for hour $t$ {(FIG. \ref{fig:data}(b))}. Unless otherwise stated, \textbf{we only use the normalized average fluorescence data} (FIG. \ref{fig:data}(b)) from this point forward. In addition, we refer to the ``fluorescence data'' simply as ``data''.
	
	We use the initial data $\tilde V_0(r)$ {(cells not subject to nocodazole)} as the initial condition {and the final data $\tilde V_{24}(r)$ (cells after 24h of nocodazole exposure) as the steady state solution} for our partial differential equation {(PDE)} models, which require smooth functions on $\mathbb{R}$. A straightforward choice is to approximate the initial data using a sum of Gaussians. An approximation is shown in FIG. \ref{fig:data}(b), inset, where the average initial data (gray) is plotted with its approximation (dashed green) superimposed.  Generally, if $\tilde V_0(r)$ {or $\tilde V_{24}(r)$} appear in model equations, {they} represent the Gaussian approximations and otherwise represent data.
	
	\subsection{Mathematical Models of Vimentin Organization}
	The primary goal of this paper is to construct a minimal model of vimentin spatial distribution in cells after triggering microtubule depolymerization. To drive our modeling efforts, we take note of salient qualitative features {over time} in the normalized data (FIG. \ref{fig:data}(b)): upon microtubule depolymerization, the vimentin profile advects from the cell edge towards the cell center and accumulates {near the nuclear envelope} before eventually stabilizing.  
	
	To model the observed motion and stabilization, we rely on two primary mechanisms. The first mechanism involves an inward motion of vimentin from the cell edge to cell center that we name retrograde flow. This retrograde flow results from \yp{complex} actin-related dynamics, \yp{which has not been investigated experimentally} in this study. In particular, how retrograde flow depends on location within the cell or how much it transports is unknown. In the model derivation to follow, we allow the retrograde velocity to be an arbitrary function, $\ub(\cdot)$, then consider several plausible forms of $\ub(\cdot)$. 
	The second mechanism allows vimentin to switch/transition between two states; mobile vimentin and immobile vimentin. Mobile vimentin is subject to retrograde flow and thus move towards the cell center, while immobile vimentin is not subject to retrograde flow. Given a small interval in space and time, some mobile vimentin may become immobile and stop moving (``trap''), or some immobile vimentin may become loose and move with the retrograde flow (``release''). We call these processes the ``\tar'' mechanism. Trapping could be a result of cross-links between filaments or with other intracellular components that prevent intermediate filament motility  \cite{minin2008intermediate,spurny2008plectin}.
	
	\yc{In the models considered, we exclude diffusion of mobile vimentin and the net growth of filaments. In interphasic cells, the majority of intermediate filament material is assembled in an insoluble pool at any given time. More than 80\% of proteins are assembled in filaments or networks and form the insoluble pool, which is only visible with the type of microscopy considered in our experimental work -- the soluble pool may contribute only slightly to the fluorescence. Considering the diffusion of a soluble pool to describe the dynamics of the mobile vimentin would increase the complexity of the model by adding a parameter, which would be penalized in the model selection process. Furthermore, previous studies have verified that filament polymerization is negligible at the time scale of hours in primary astrocytes \cite{hookway2015microtubule,leduc2017regulation}. Protein degradation and \textit{de novo} synthesis are neglected.}
	
	\paragraph{Modeling Framework:}
	
	Let $V(\x,t)$ represent the average vimentin intensity data on the annular domain
	\begin{equation*}
		\x \in\Omega := \{(x,y): L_0 \leq \|\x\|\leq L \},
	\end{equation*}
	where $L_0$ corresponds to the nuclear envelope and $L$ corresponds to the cell edge. In our modeling framework, vimentin refers to the total fluorescent vimentin composed of soluble forms, assembled in short or long filaments and integrated in networks.
	
	We assume that the data consists of some underlying combination of \ya{immobile and mobile} vimentin: $V(\x,t) = I(\x,t) + M(\x,t)$, where $I$ represents immobile vimentin and $M$ represents mobile vimentin. Then {we model the dynamics of cellular vimentin distribution using an advection equation}
	\begin{equation*}
		\frac{\pa V}{\pa t}(\x,t) =  \frac{\pa I}{\pa t}(\x,t) + \frac{\pa M}{\pa t}(\x,t) = \nabla \cdot (\ub(\cdot) M),
	\end{equation*}
	since $I$ does not advect. {The function $\ub(\cdot)$ represents retrograde flow {velocity} and is a vector-valued function where its output vector points towards the origin.} We define the equation for $I$ to be,
	\begin{equation}\label{eq:t}
		\frac{\pa I}{\pa t} = \underbrace{\vphantom{\beta}\alpha M }_\textrm{Trap} - \underbrace{\beta I}_\textrm{Release},
	\end{equation}
	where {the right-hand side is the} \tar mechanism defined by a first-order exchange between mobile and immobile vimentin. The parameters $\alpha$ and $\beta$ are the trap rate and release rate, respectively. Then the model equations are
	\begin{equation}\label{eq:2d}
		\begin{split}
			\frac{\pa I}{\pa t} &= \alpha M  - \beta I,\\
			\frac{\pa M}{\pa t} &= \nabla \cdot (\ub(\cdot) M) - (\alpha M - \beta I).
		\end{split}
	\end{equation}
	We assume that $V$ is circularly symmetric in agreement with the data. While {$I$ and $M$, the component parts of $V$, are plausibly not circularly symmetric, we also assume circular symmetry in $I$ and $M$ for simplicity}. Hence, we exploit circular symmetry of the underlying solutions and transform \eqref{eq:2d} to polar coordinates, where we discard the angular coordinate and restrict the domain to a one-dimensional radial line on the interval $[L_0, L]$. {Thus, \eqref{eq:2d} becomes}
	\begin{equation}\label{eq:model}
		\begin{split}
			\frac{\pa I}{\pa t}(r,t) &= \alpha M  - \beta I,\\
			\frac{\pa M}{\pa t}(r,t) &= \underbrace{\frac{1}{r} \frac{\pa}{\pa r}(r u(\cdot) M(r,t))}_\textrm{Advection/Retrograde Flow} - \underbrace{\vphantom{\frac{\pa 1}{\pa 1}}(\alpha M - \beta I)}_\textrm{{Trap \& Release}},
		\end{split}
	\end{equation}
	{considered with the initial conditions,
		\begin{equation*}\label{eq:ic}
			I(r,0) = \ve\tilde V_\text{0}(r), \quad M(r,0) = (1-\ve)\tilde V_\text{0}(r),
		\end{equation*}
		where {the free parameter} $\ve\in[0,1]$ represents the initial proportion of immobile vimentin in the data $\tilde V_\text{0}(r)$ (FIG. \ref{fig:data}(b), inset).} {While the data show a residual quantity of vimentin at the cell edge $L$ at some times, we observe a decay in vimentin at the cell edge over time}. For simplicity, we choose a homogeneous Dirichlet condition at the cell edge,
	\begin{equation*}
		M(L,t) = 0,
	\end{equation*}
	{for all $t\geq 0$.}
	
	\paragraph{{Assumptions and Hypotheses:}}
	Using our modeling framework defined in \eqref{eq:model}, we now explore our model assumptions {with} distinct {assumptions and} hypotheses {and formulate different scenarios}.
	
\begin{figure}[t]
		\centering
		\includegraphics[width=\ww\textwidth]{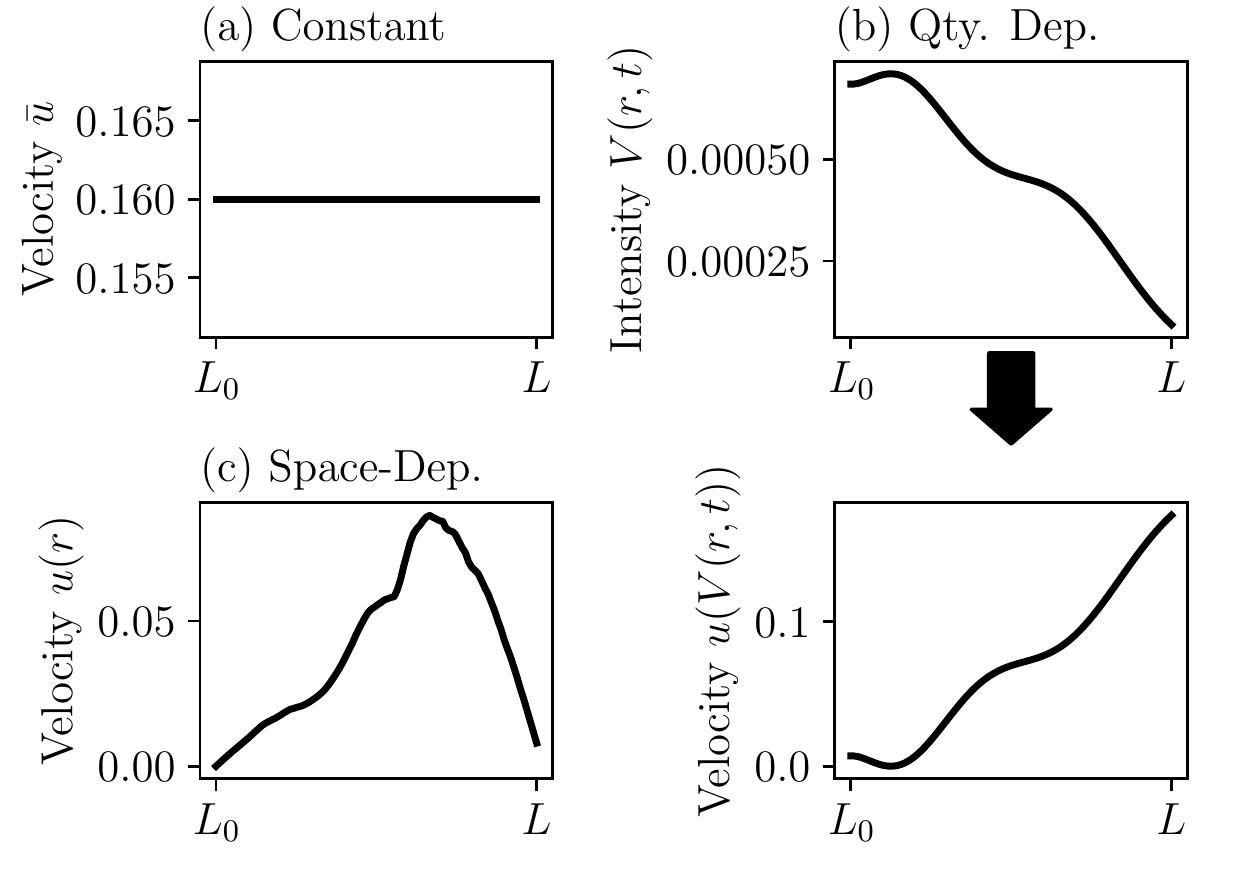}
		\caption{Example velocity profiles $u(\cdot)$. (a): An example of a constant velocity profile, $\bar{u}=\SI{0.16}{\um/\min}$ \cite{leduc2017regulation}. (b), top: An example solution of model \eqref{eq:model}, where $V(r,t) = I(r,t) + M(r,t)$, at $t=\SI{1}{h}$ with $\ve=1$, $\alpha=\beta=0$. (b), bottom: A plot of $u(V(r,t))$, the corresponding vimentin quantity-dependent velocity defined by \eqref{eq:u_conc} with $V_\text{m}=\num{7e-4}$, $u_\text{m}=0.2$ and $V(r,t)$ is taken from (b), top. (c): An example of a spatially-dependent velocity \eqref{eq:u}.}
		\label{fig:velocity}
	\end{figure}

	{\bf All Constant Parameters - T1 -} Here, {we assume} velocity is a constant, free parameter and denoted by $u(\cdot) = \bar u$ (FIG. \ref{fig:velocity}(a)). The remaining free parameters $\ve\in[0,1]$ and $\alpha,\beta\geq0$ are also constant parameters.
	
	{\bf Quantity-Dependent Velocity - T2 -} We assume that $u(V)$ is a monotonically decreasing function of {vimentin quantity} $V$ (i.e., $d u/d V < 0$). This mechanism is phenomenologically described by the simplest linear choice for $u(V)$:
	\begin{equation}\label{eq:u_conc}
		u(V) = u_\text{m}\left(1 - \frac{V}{V_\text{m}}\right),
	\end{equation}
	where $u_\text{m}$ and $V_\text{m}$ are free parameters and represent the maximum retrograde velocity and maximum vimentin quantity, respectively. An example of this function is shown in FIG. \ref{fig:velocity}(b), bottom, given an example solution $V(x,t)$ at $t=\SI{1}{h}$: the velocity is relatively small near maximum vimentin quantity, whereas the velocity is relatively large in regions of lower vimentin quantity. Velocity reduces as a function of vimentin quantity, resulting in an effect similar to a traffic jam. 
	
	For {the assumptions} with constant parameters (T1) and quantity-dependent velocity (T2), we consider the following {hypotheses}:
	\begin{enumerate}[label=\Alph*.,ref=\Alph*]
		\item \label{hypothesis:t_and_r} Trap and release: mobile vimentin may become immobile and vice-versa. $\alpha,\beta>0$.
		\item \label{hypothesis:release} Only release, no trapping: only the {initial} proportion of immobile vimentin can become mobile and mobile vimentin can not become immobile. $\beta>0$, $\alpha=0$.
		\item \label{hypothesis:transport} Pure transport with neither trap nor release: mobile vimentin do not become immobile and vice-versa. $\alpha=\beta=0$.
		\item \label{hypothesis:trap}Only trap, no release \yp{(irreversible trapping)}: mobile vimentin can only become immobile and immobile vimentin can not become mobile. $\alpha>0$, $\beta=0$.
	\end{enumerate}
	
	\yp{We remark that assumptions} T1 and T2 with nonzero velocity do not obey conservation of mass because some mobile vimentin is guaranteed to leave the domain. However, we are allowed to assume conservation of mass for \yp{assumption} T3 below.
	
	
	{\bf Spatially-Dependent Terms - T3 -} Here, we {assume some model terms to depend on space and consider the following hypotheses}:
	
	
	\begin{enumerate}[label=\Alph*.,ref=\Alph*,start=5]
		\item \label{hypothesis:u} Irreversible trapping and spatially-dependent velocity depicting cell compartmentalization \yp{with conservation of mass}: Suppose that $\beta = 0$ and let $\alpha \geq 0$ be constant. Then it is possible to write down the spatially-dependent velocity \ya{by assuming that the steady-state distribution of vimentin is immobile and equal to the experimental data at 24 hours $V^*(r)=I^*(r) := \tilde V_{24}(r)$} \ya{(see Appendix \ref{a:u} for a detailed derivation):}
		\begin{equation}\label{eq:u}
			u(r) = \frac{\alpha}{r\hat I(r)} \int_{L_0}^r s  \left(\hat  I(s) + M_0(s)\right) \,\mathrm{d}s,
		\end{equation}
		where $\hat I(r) := I_0(r) - I^*(r)$, $I^*(r) := \tilde V_{24}(r)$, $I_0(r) := I(r,0)=\ve\tilde V_\text{0}(r)$ and $M_0(r) := M(r,0)=(1-\ve)\tilde V_\text{0}(r)$. The parameters $\alpha>0$ and $\ve\in[0,1]$ are the only two free parameters. An example of this function is shown in FIG. \ref{fig:velocity}(c). 
		\item \label{hypothesis:alpha} Spatially-dependent net attachment rate and constant velocity \yp{with conservation of mass}: Similarly, if we set $\beta=0$, assume a constant velocity $u(\cdot) = \bar u$, \ya{and assume that the steady-state distribution of vimentin is immobile and equal to $\tilde V_{24}(r)$}, then it is possible to write down the spatially-dependent rate,
		\begin{equation}\label{eq:alpha}
			\alpha(r) = \frac{r\bar{u}\hat I(r)}{\int_{L_0}^r s(\hat I(s) + M_0(s)) \,\mathrm{d}s}.
		\end{equation}
		Note that we allow this function to be negative, and thus represents a \textit{net} exchange between mobile and immobile vimentin. Hence, $\alpha(r)$ represents the spatially-dependent net attachment rate. The parameters $\bar{u}>0$ and $\ve\in[0,1]$ are the only two free parameters. 
	\end{enumerate}
	
	In these spatially-dependent scenarios T3E and T3F, the retrograde flow velocity and trapping rate are {closely related}: $\alpha$ is the main determinant of the speed for T3E and $\bar u$ scales the magnitude of the net attachment rate $\alpha(r)$ for T3F.
	
	\begin{table}[t]
		\caption{(a) Description of free parameters with units.  (b) parameter dependency for each hypothesis. Parameters with dimension 1 are dimensionless. Check marks (\ckb) denote free parameters to be {estimated}, while \xx marks denote parameters that are to be excluded for a given {hypothesis}. We {implement different hypotheses} by setting corresponding parameters to zero. Purple cells correspond to the {all constant parameters} (T1), light brown cells correspond to the {quantity-dependent velocity assumption $u(V)$} (T2), and gray cells correspond to the spatially-dependent {assumption}, $u(r)$ or $\alpha(r)$ (T3).}\label{tab:free_pars}
		\subfloat[\label{A}][]{
			\begin{tabular}{p{0.15\textwidth}<{\centering}|p{0.55\textwidth}|p{0.15\textwidth}}
				Param. & Description & Units\\
				\hline
				$\ve$ & Initial proportion of immobile vimentin & 1\\
				\hline
				$\alpha$ & Trapping rate& \si{1/\min}\\
				\hline
				$\beta$ & Release rate&\si{1/\min}\\
				\hline
				$\bar u$ & Constant retrograde velocity & \si{\um/\min}\\
				\hline
				$u_\text{m}$ & Maximum velocity in the jamming mechanism \eqref{eq:u_conc} & \si{\um/\min}\\
				\hline
				$V_\text{m}$ & Maximum {quantity} $V$ in the jamming mechanism \eqref{eq:u_conc}& 1
		\end{tabular}}
		\setlength\tabcolsep{.006in}
		\subfloat[\label{B}][]{
			\centering
			\begin{tabular}{l|cccc|ccccc|cccc}
				& \multicolumn{4}{c|}{Const. (T1)}       & \multicolumn{5}{c|}{Qty. (T2)}                                 & \multicolumn{4}{c}{Space (T3)}               \\ \cline{2-14} 
				{Hypothesis}                  & $\varepsilon$ & $\alpha$ & $\beta$ & $\bar{u}$ & $\varepsilon$ & $\alpha$ & $\beta$ &  $u_\text{m}$ & $V_\text{m}$ & $\varepsilon$ & $\alpha$ & $\beta$ & $\bar{u}$ \\
				\hline
				A. Trap \& Rel.      &\ckp        &\ckp    &\ckp    &\ckp  &\ckb  &\ckb     &\ckb     &\ckb    &\ckb   & & & & \\
				B. Release           &\ckp        &\bkp\xx &\ckp    &\ckp  &\ckb  &\bkg\xx  &\ckb     &\ckb    &\ckb   & & & \\
				C. Transport         &\ckp        &\bkp\xx &\bkp\xx &\ckp  &\ckb  &\bkg\xx  &\bkg\xx  &\ckb    &\ckb   & & & & \\
				D. Trap              &\ckp        &\ckp    &\bkp\xx &\ckp  &\ckb  &\ckb     &\bkg\xx  &\ckb    &\ckb   & & & & \\
				\hline
				E. Trap, $u(r)$      &               &          &         &         &               &                &              &        &       &\cko        &\cko          &\bko\xx       &\bko\xx         \\
				F. Trap, $\alpha(r)$ &               &          &         &         &               &                &              &        &       &\cko        &\bko\xx          &\bko\xx       & \cko      
		\end{tabular}}
	\end{table}
	
	\begin{figure*}
		\centering
		\includegraphics[width=.9\textwidth]{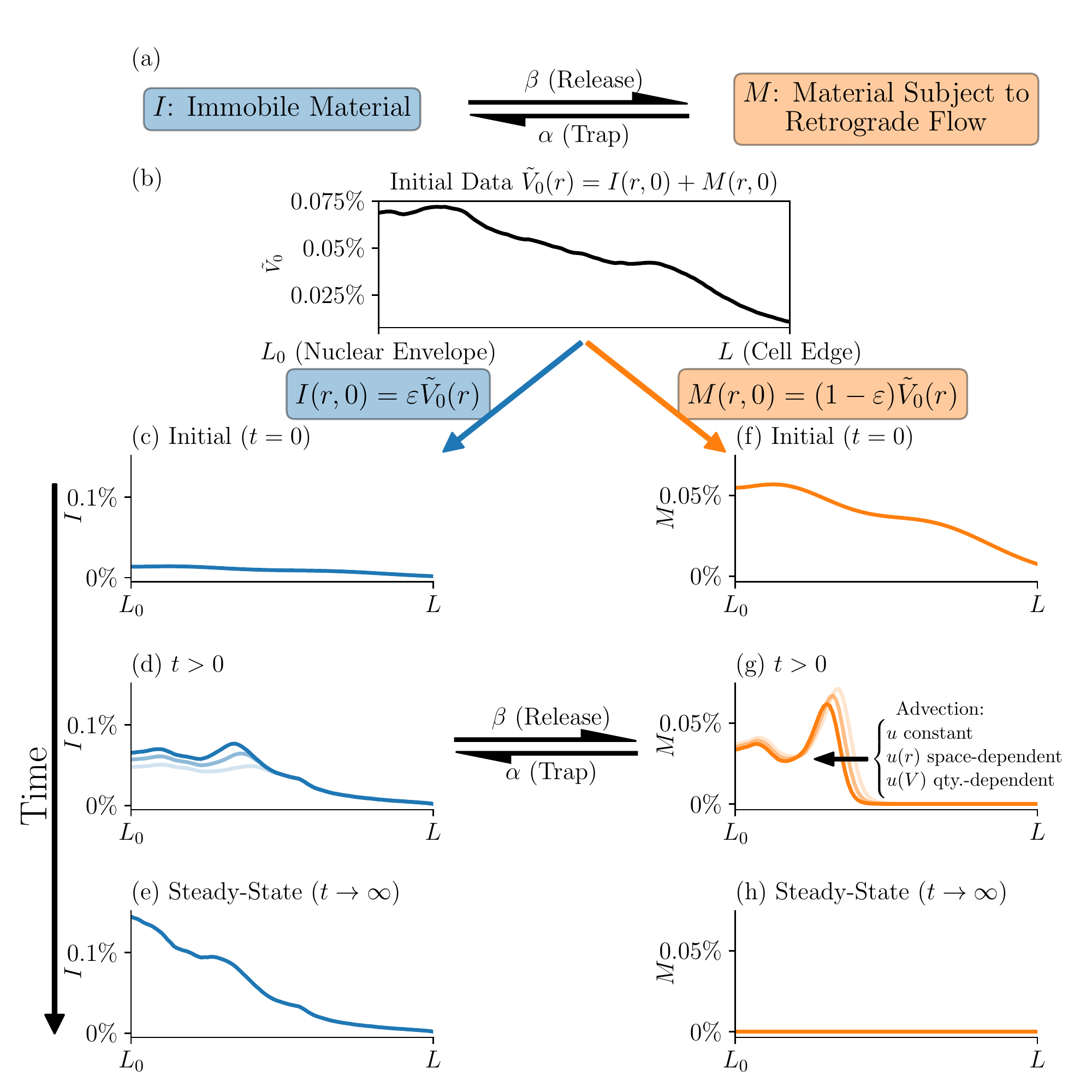}
		\caption{Mechanisms considered in the modeling framework \eqref{eq:model}. \yp{For illustration, we show solutions of scenario T3E (irreversible trapping with spatially-dependent retrograde velocity).} (a): The model involves two populations of vimentin: immobile ($I$) and mobile ($M$). The mobile vimentin is subject to retrograde flow, while the immobile vimentin is stationary. (b): Initial data. The data is assumed to be the sum of mobile and immobile vimentin, i.e., $V(r,t) = I(r,t) + M(r,t)$. The initial conditions for $I$ and $M$ are taken to be some proportion of the initial data\yp{, $\tilde V_\text{0}(r)$,} determined by the parameter $\ve$ ((c), (f)). (d), (g): The model uses two mechanisms. The first mechanism involves advection of mobile vimentin towards the left boundary $L_0$ (panel (g)). The advection velocity may be taken to be constant, space-dependent, or quantity-dependent. \yp{If advection \yp{velocity} is constant or quantity-dependent, it is possible for mobile material to advect out of the domain (hypotheses T1 and T2, resp.)}. The second mechanism involves a transition rate from mobile (immobile) to immobile (mobile) denoted by $\alpha$ ($\beta$). (e), (h): Steady-state. \yp{After enough time, all mobile material have become immobile (panel (e))}. Parameters: $\ve=0.2$, $\alpha$=0.01, $\beta=0$.}
		\label{fig:solution_schematic}
	\end{figure*}
	
	We illustrate our modeling framework \eqref{eq:model} using scenario T3E (irreversible trapping and spatially-dependent retrograde flow $u(r)$) in FIG. \ref{fig:solution_schematic}. Descriptions of parameters are in TABLE \ref{tab:free_pars}(a) with a summary of free parameters for each hypothesis in TABLE \ref{tab:free_pars}(b). 
	For convenience, we let $\mathcal{V}_s:=\{\text{T1A},\text{T1B},\text{T1C},\text{T1D},\text{T2A},\text{T2B},\text{T2C},\text{T2D},\text{T3E},\allowbreak\text{T3F}\}$ denote the index set of scenarios considered in this work.

	\subsection{Model Evaluation}
	Recall \ya{that we use the notation} $\mathcal{V}_t$, $\mathcal{V}_r$, and $\mathcal{V}_s$, \ya{for the} index \ya{set} of observation times, index \ya{set} of observation positions, and the index \ya{set} of scenarios, respectively. Given a scenario $i\in {\mathcal{V}_s}$ and corresponding parameters $\p_i$ {(Table \ref{tab:free_pars}(b))}, {we numerically evaluate the {scenario} using method of lines with \ya{a forward Euler scheme} (Appendix \ref{a:numerical_scheme}). We then compare the solution against the data by using the residual sum of squares} (RSS)
	\begin{equation}\label{eq:rss}
		\rss_i(\p_i) = \sum_{t\in {\mathcal{V}_t}} \sum_{r \in {\mathcal{V}_r}} \left(\tilde{V}_t(r) - V_i(\yp{r,t},\p_i)\right)^2,
	\end{equation}
	
	\noindent where \yp{$\tilde{V}_t(r)$ is the experimental normalized vimentin quantity and} $V_i(\yp{r,t},\p_i)$ is the $i$th {scenario}'s \yp{solution} (given by the sum of mobile and immobile vimentin, $V_i=I_i+M_i$). The vimentin profiles are relatively small in magnitude, on the order of $\num{1e-4}$, so the $\rss_i$ values are relatively small for our problem. To make parameter space minima more apparent, we use the {base-10} log of the error during the optimization procedure. Furthermore, {since experimental vimentin profiles approach a steady state \yp{(FIG. \ref{fig:data}(b))},} we impose a condition to ensure that the solution reaches a steady state within the \SI{24}{h} experimental time frame. Hence, the optimization error is expressed:
	\begin{equation}\label{eq:opt}
		{\Delta}(\p_i) = 
		\begin{cases}
			\log_{10}\left(\rss_i(\p_i) \right)& \text{if steady-state}\\
			10^5 & \text{else}\\
		\end{cases}.
	\end{equation}
	We define the steady-state condition to be true when
	\begin{equation*}
		\sum_{r} \left(V(\yp{r,t_1},\p_i) - V(\yp{r,t_2},\p_i)\right)^2< \delta,
	\end{equation*}
	where ${\delta}=\num{1e-10}$, $t_1 = \SI{20}{h}$, and $t_2=\SI{24}{h}$. The steady-state condition forces the error to be a relatively large value if the model solution is not at steady-state \textit{towards the end of the simulation}. In other words, if the PDE solution at \SI{20}{h} and \SI{24}{h} differs by more than the threshold amount $\delta$, then we say that the solution has not reached steady-state and we return a relatively large error of $10^5$. Hence parameter sets for which the \yp{scenario} solution does not reach steady state {after \SI{20}{h}} are disqualified.
	
	For each {scenario} $i \in \mathcal{V}_s$, we minimize \eqref{eq:opt} over the space of parameters and take the exponent to recover the \rss:
	\ya{\begin{align*}
			\rss_i(\p_i^*)&=\min_{\p_i} \left[\rss_i(\p_i)\right]= 10^{\left(\min_{\p_i} \left[{\Delta}(\p_i)\right]\right)},
	\end{align*}}
	where $\p_i^*$ is the estimate of $\p_i$ that yields the minimal error. We refer to $\p_i^*$ as the optimal parameter set and use asterisks on individual parameters if they are part of an optimized parameter set. We use the differential evolution function available in Python's scipy package \cite{2020SciPy-NMeth} and run the differential evolution 100 times for each {scenario} using a tolerance of \num{1e-4}.
	
	We discriminate between the ten {($=|\mathcal{V}_s)|)$} scenarios in this study using the model selection Akaike information criterion (see for details Appendix \ref{sec:model_selection}). {Then}, further investigations are carried out on two biologically plausible scenarios that yield the best model outputs. Confidence intervals for parameters values are computed using the log-likelihood ratio statistic (see Appendix \ref{a:confidence_interval} for details). {Finally,} a global sensitivity analysis with eFAST is then carried out to determine the driving parameter(s) of {these two} scenarios. We use the optimal parameter values as baseline values from which we perturb in the space of parameters, then quantify the resulting change in the \rss \eqref{eq:rss}. The \rss is most likely non-monotonic and nonlinear for both models so we use eFAST \cite{marino2008methodology}, which is appropriate in this case and returns two key quantities: the first-order sensitivity index $S_{k}$ for parameter $k$, which simply measures the variance in the the \rss (and thus the model output) as a result of variance in parameter $k$, and the total-order sensitivity index $S_{T_{k}}$, which measures the sum all of first and higher-order interactions between the parameters.
	
	\section{Results}
	
	\begin{figure*}
		\centering
		\includegraphics[width=\textwidth]{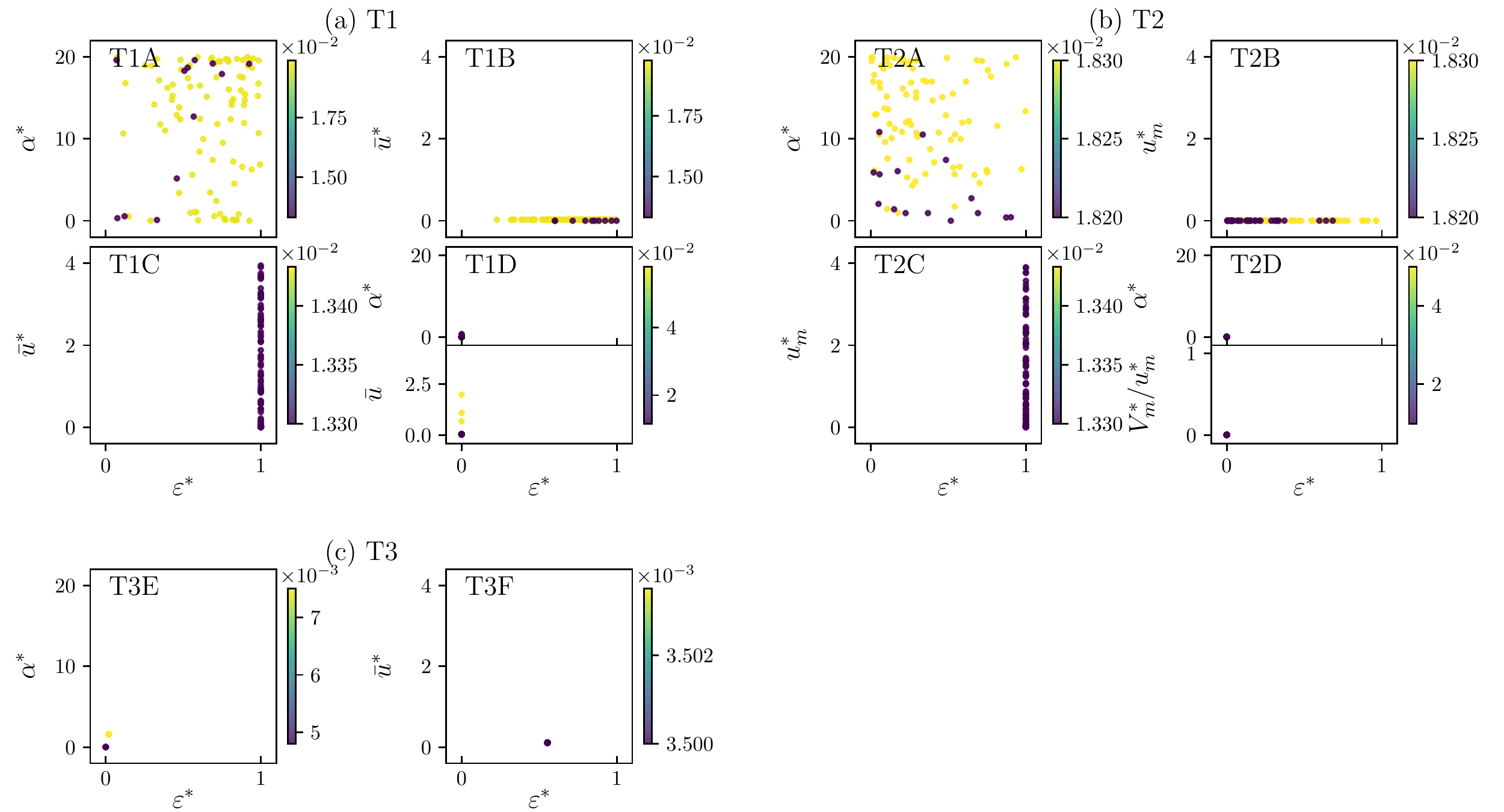}
		\caption{Scatter plots of parameter estimates. {For each scenario, we obtain 100 parameter estimates by minimizing the \rss with differential evolution}. Darker shades correspond to lesser \rss values and lighter shades correspond to greater \rss values. Parameter values with darker shades are considered optimal parameter values. The clustering/spatial distribution of optimal parameters in the parameter spaces informs on the parameter identifiability for each scenario. Hence, scenarios T1D, T2D, T3E, and T3F are identifiable. All \rss values are rounded to the fourth decimal place.}
		\label{fig:identifiability}
	\end{figure*}
	
	The ten {scenarios} of the collection $\mathcal{V}_s$ are calibrated using the same average vimentin data (FIG. \ref{fig:data}(b)). For a given {scenario}, we perform the calibration by searching for parameter values resulting in solutions that most closely reproduce the data (i.e, by minimizing the optimization error \eqref{eq:opt} using differential evolution \cite{storn1997differential}). We repeat this calibration 100 times, yielding a collection of up to 100 optimal parameter sets. {For the reader's convenience, we include solutions of 10 parameter sets with the lowest \rss (out of the 100 parameter sets) of all scenarios in FIG. \ref{fig:sols}.}
	
	\subsection{Model Identifiability}
	For each {scenario}, we plot parameter value-pairs found by the 100 runs of the optimization in FIG. \ref{fig:identifiability}. For a {scenario} to be considered identifiable, the parameters values providing the global minimum of the RSS must be tightly ``clustered'' in parameter space (intuitively, identifiable scenarios are ``good models'' \yp{as they} have single values \yc{or finite confidence intervals} for parameters \yp{when calibrated to data}). Only scenarios T1D, T2D, T3E, and T3F satisfy these requirements. For each of these scenarios, the global minimum, marked by dark dot(s), are tightly clustered. In contrast, the remaining {scenarios}, those under {hypotheses} \ref{hypothesis:t_and_r} through \ref{hypothesis:transport} exhibit a relatively great degree of dispersal.
	
	{In hypothesis \ref{hypothesis:t_and_r}, which results in the most complex scenarios, none of the parameters are identifiable.} In {hypothesis} \ref{hypothesis:release}, where {scenarios} only possess the release mechanism with no trapping ($\alpha=0$ and $\beta>0$), the optimal velocity for the retrograde flow is found to be $\bar{u}^*=u_\text{m}^*=0$ {and the initial proportion of immobile vimentin $\ve^*$ is not identifiable} (FIG. \ref{fig:identifiability} T1B, T2B). Hence, if all vimentin eventually become mobile and eventually no immobile vimentin exists, the only way to recover the data is to assume no transport, which is inconsistent with the working biological assumption. Furthermore, FIG. \ref{fig:identifiability} T1B, T2B show that forcing the velocity to be zero results in \yp{practically} non-identifiable scenarios. For all scenarios under hypothesis \ref{hypothesis:transport}, i.e., scenarios with pure transport ($\alpha=\beta=0$), the velocity parameter $\bar{u}$ is not identifiable\ya{. T}he calibration results in either zero velocity $\bar{u}^*=u_\text{m}^*=0$ (which violates the biological assumption and can not be considered), or a positive velocity ($\bar{u}^*,u_\text{m}^*>0$). In {the latter} case, the optimization finds $\epsilon^*=1$ (FIG. \ref{fig:identifiability} T1C, T2C), so the {scenarios} initialize with only immobile vimentin; because there is no exchange between mobile and immobile, the initial immobile vimentin is the solution for all time.
	
	In summary, \ya{scenarios with no decay of the mobile part ($\alpha=0$, B and C) are non-identifiable and parameter searches for these scenarios find that vimentin must be purely immobile to describe the data.} Furthermore, only the identifiable {hypotheses} (D, E, and F) allow mobile vimentin to become immobile with no transition from immobile to mobile by definition ($\beta=0$). Hence, we conclude that the existence of immobile vimentin is necessary to explain the data.
	
	\subsection{Model selection results}
	
	\setlength\tabcolsep{.01in}
	\begin{table}
		\caption{Akaike information criterion {($\text{AIC}_i$) values for each scenario $i\in \mathcal{V}_s$} and Akaike weights {$w_i$}. {The parameter $K_i$ indicates the number of parameters estimated for the computation of $\text{AIC}_i$ values.}}\label{tab:aic}
		\centering
		\makebox[\textwidth][c]{
			\begin{tabular}{l|cccl|cccl|ccc}
				&\multicolumn{3}{c}{T1} & & \multicolumn{3}{c}{T2} & & \multicolumn{3}{c}{T3}\\
				\cline{2-4} \cline{6-8} \cline{10-12}
				{Hyp.} &  $K_i$ & $\text{AIC}_i$ & $w_i$ &  & $K_i$ & $\text{AIC}_i$ & $w_i$ &  & $K_i$ & $\text{AIC}_i$ & $w_i$\\ 
				\hline
				A.   &\bkp5&\bkp-16861.48&\bkp0 && \bkg6&\bkg-16137.10&\bkg0 && & &\\
				B. &\bkp4&\bkp-16863.48  &\bkp0 && \bkg5&\bkg-16144.27&\bkg0 && & &\\
				C.  &\bkp3&\bkp-16865.12 &\bkp0 && \bkg4&\bkg-16863.12&\bkg0 && & &\\
				D. &\bkp4&\bkp-17204.23  &\bkp0 && \bkg5&\bkg-17541.09&\bkg0 && & &\\
				\hline
				E. &     &             &      &&      &             &      &&\bko3&\bko-19236.60&\bko0\\
				F.&     &             &      &&      &             &      &&\bko3&\bko-19862.73&\bko1\\
		\end{tabular}}
	\end{table}
	
	
	We include all identifiable and non-identifiable {scenarios} in the model selection procedure because it is possible for a non-identifiable {scenario} to outperform other {scenarios} under AIC (only the minimal \rss is needed -- parameter {properties and values} are not considered). {It would be undesirable for model selection to choose a non-identifiable scenario, however, choosing an identifiable scenario over non-identifiable scenarios would provide an additional degree of confidence in the chosen scenario}.
	
\begin{figure}[ht!]
		\centering
		\includegraphics[width=.75\textwidth]{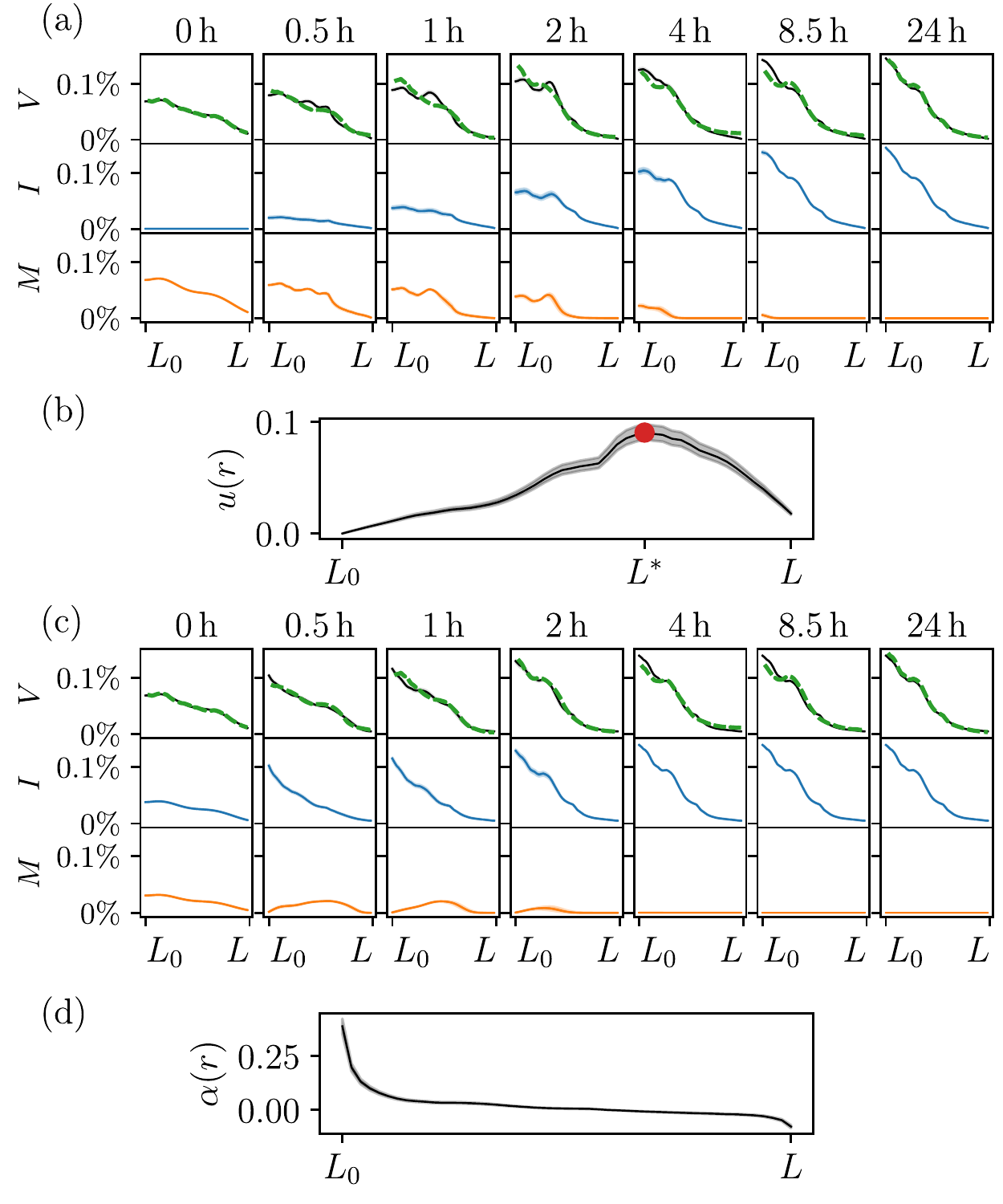}
		\caption{Solutions of scenarios {T3E and T3F}. Each column in (a), (c) corresponds to an experimental time point $t \in \mathcal{V}_t$. Black curves correspond to the PDE solution for total vimentin, $V=I+M$ with experimental data overlaid in dashed green. Blue curves correspond to the PDE solution for immobile vimentin $I$ and orange curves correspond to the PDE solution for mobile vimentin $M$. Shaded regions denote \yp{solutions obtained with parameter values in} confidence intervals in the most sensitive parameter (see FIG. \ref{fig:cost_functions}(b), (d)) {(a): Scenario T3E, spatially-dependent retrograde velocity $u(r)$ with constant trapping rate $\alpha$. Optimal parameter values are $\ve^*=0$ and $\alpha^*=0.011$. (b): Retrograde velocity \yp{profile}  \eqref{eq:u} as a function of space. (c): Scenario T3F, spatially-dependent net trapping rate $\alpha(r)$ with constant velocity. Optimal parameter values are $\ve^*=0.55$ and $\bar{u}^*=0.11$. (d): Net trapping rate function \eqref{eq:alpha} as a function of space. For the optimized solutions of all other scenarios, see Appendix \ref{a:solutions}}.}\label{fig:best2sol}
	\end{figure}

	AIC values provide a means to rank the different {scenarios}. Akaike Information Criterion (AIC) values and Akaike weights for each {scenario} are shown in TABLE \ref{tab:aic}. Note that {scenarios} with the greatest AIC values happen to be the non-identifiable hypotheses \ref{hypothesis:t_and_r} to \ref{hypothesis:transport} (the top-ranked {scenario} has the lowest AIC). To make a conclusive determination of the best {scenario}, we use the Akaike weights $w_i$, which tells us that \textbf{scenario T3F, {which is characterized by a spatially-dependent net trapping rate}} \yp{(FIG. \ref{fig:best2sol}(c))}, is an unambiguous choice to best represent the experimental data among the collection of \yp{considered} scenarios.
	
	Going further in the interpretation of the model selection results, we note that {with} no accumulation of immobile vimentin via trapping (hypotheses \ref{hypothesis:release} and \ref{hypothesis:transport} where $\alpha=0$), the corresponding {scenarios} are ranked poorly, reinforcing the previous observation that \yp{vimentin trapping is required to explain the experimental observations}. In particular, note that {scenarios T1C and T1B, which differ in the parameter $\beta>0$ (release rate),} obtain close AIC values, making the model selection method inconclusive for discriminating between these scenarios. Next, {when only considering assumption T1 (resp. T2), hypothesis \ref{hypothesis:t_and_r} (which has the greatest number of parameters) ranks the lowest, whereas hypothesis \ref{hypothesis:trap} always ranks the highest}. Indeed, scenario T2D is the third best scenario and is characterized by $\beta=0$ with a crowding effect described by a quantity-dependent velocity. 
	
	Again considering all scenarios, the poor ranking of the constant-velocity and quantity-dependent velocity {assumptions} relative to the spatially-dependent {assumptions, T3E and T3F,} is consistent with the biological observations of spatially-dependent retrograde flow of actin \cite{jiu2015bidirectional} and the underlying compartmentalization of cell organelles interacting with vimentin \cite{gao2001novel,toivola2005cellular,chang2009dynamic,nekrasova2011vimentin,cremer2022vimentin}, respectively. Because both scenarios are biologically plausible, we examine them in {more} detail.
	
	{\subsection{Best Scenarios: Spatially-Dependent Assumption}}
	
	{Solutions of both scenarios T3E and T3F compared to data are shown in FIG. \ref{fig:best2sol}(a) and \ref{fig:best2sol}(c) along the top rows. The initial and final solutions are expected to match exactly because of how we define the initial condition and how we derive the spatially-dependent velocity \eqref{eq:u} for T3E and spatially-dependent trapping rate \eqref{eq:alpha} for T3F (in particular, defining the steady-state solution of the PDE to match the data at \SI{24}{h} is a key assumption \yp{(see Appendix \ref{a:spatially-dependent})}). Thus, the contribution lies in \yp{re}covering the solutions at intermediate times and the strong agreement between the model solution and the data at these intermediate times.}
	
	\paragraph{Model Robustness:} We now determine the robustness of scenarios {T3E and T3F} to perturbations in their optimized parameter pairs, $(\ve^*,\alpha^*)$ and  $(\ve^*,\ya{\bar{u}^*})$, respectively. We use the respective optimal values {$(\ve^*,\alpha^*)=(0,0.11)$ and $(\ve^*,\bar{u}\yp{^*})=(0.55,0.11)$} as baseline values from which we perturb in the space of parameters, then quantify the resulting change in the \rss \eqref{eq:rss}. The results of the {sensitivity} analysis are shown in TABLE \ref{tab:sensitivity}.
	
	\renewcommand{\arraystretch}{1.2}
	\begin{table}[t]
		\caption{\yp{Scenario} {T3E} (a) and {T3F} (b) parameter sensitivity determined using eFAST. \ya{$S_k$} is the first-order sensitivity index of parameter \yp{$k$} and \yp{$S_{T_k}$} is the total sensitivity of parameter \yp{$k$}. \yp{Baseline parameter values correspond to the optimal parameter value(s), and the range denotes the parameter range over which the sensitivity analysis is performed.}} \label{tab:sensitivity}
		\setlength\tabcolsep{.04in}
		\centering
		\subfloat[\label{A}][T3E]{
			\begin{tabular}{cllll}
				Par.& \ya{$S_k$} & \ya{$S_{T_k}$} & Range & Baseline\\ 
				\hline
				$\ve$ & ${\num{3.6e-6}}$&${\num{5.3e-3}}$&$[0,5.\num{0e-2}]$&$ 0.0$\\
				$\alpha$ &0.88&0.97&$[0,\num{5.0e-2}]$& 0.011
		\end{tabular}}
		\setlength\tabcolsep{.05in}
		\subfloat[\label{B}][T3F]{
			\centering
			\begin{tabular}{cllll}
				Par.& \ya{$S_k$} & \ya{$S_{T_k}$} & Range & Baseline\\ 
				\hline
				$\ve$ & ${\num{6.6e-3}}$&${\num{4.6e-2}}$&$[0,0.7]$&$ 0.55$\\
				$\bar{u}$ &0.96 &1.00&$[0,0.3]$&$ 0.11$
		\end{tabular}}
	\end{table}

	{Scenario T3E} is least sensitive to the proportion of initial immobile vimentin, $\ve$, while virtually all model output variance is explained by parameter $\alpha$ (TABLE \ref{tab:sensitivity}(a)), which controls both the transition rate from immobile to mobile and the magnitude of the spatially-dependent retrograde velocity profile $u(r)$ \eqref{eq:u}. {Similarly, scenario T3F} is least sensitive to the proportion of initial immobile vimentin, $\ve$, while virtually all model output variance is explained by parameter $\bar{u}$ (TABLE \ref{tab:sensitivity}(b)), which determines the average retrograde velocity and the magnitude of the spatially-dependent net attachment rate.

	\begin{figure*}[ht!]
		\centering
		\includegraphics[width=.9\textwidth]{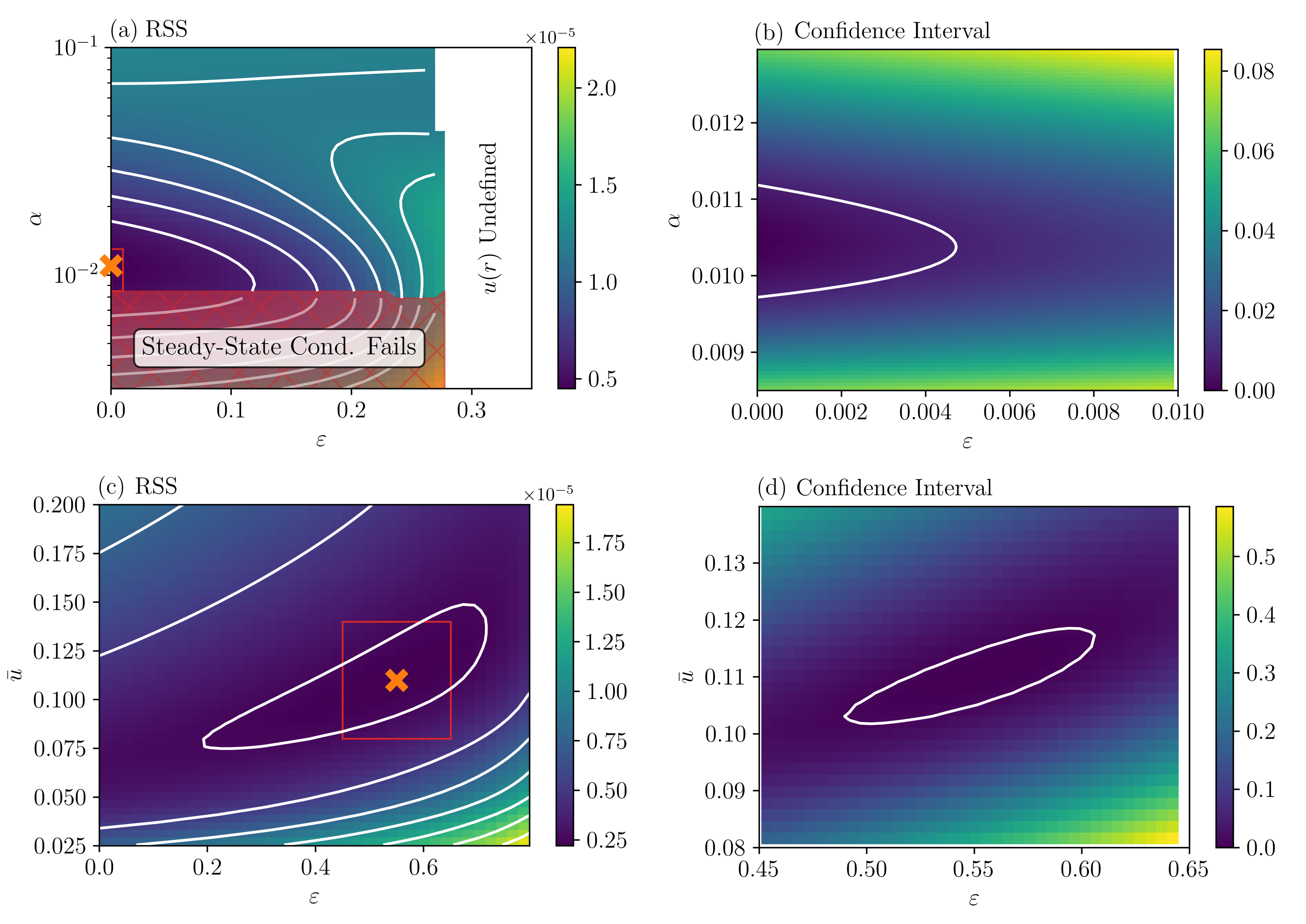}
		\caption{Visualization of the \rss \eqref{eq:rss} for scenarios T3E and T3F. As indicated by the color bars, lighter (darker) shades correspond to higher (lower) values of the \rss. The unique global minimum is denoted by an orange x at $\p^*=(\ve^*,\alpha^*)=(0,0.011)$ in (a) and at $\p^*=(\ve^*,\bar u^*)=(0.55,0.11)$ in (c). The white curve in (b), (d) delimits the confidence region with a significance level of $p=0.05$. We obtain confidence intervals using the likelihood ratio statistic \cite{raue2009structural}. Colors in (b), (d) denote the approximate $\chi^2$ values given by the difference $ \ln\left[{\rss(\bm{\theta}_0)/N}\right] - \ln\left[{\rss(\p^*)/N}\right]$, where $\bm{\theta}_0$ is an arbitrary choice of parameters \yp{$(\ve_0,\alpha_{0})$ for T3E (resp. $(\ve_0,\bar u_0)$ for T3F)} (see Appendix \ref{a:confidence_interval}). Lower $\chi^2$ values correspond to parameters with greater than 95\% confidence assuming three free parameters (\yp{the two model parameters} and variance \yp{of the error assumed in the statistical model}). (a), (b): Scenario {T3E} RSS and confidence interval, respectively. (c), (d): Scenario {T3F} RSS and confidence interval, respectively.}
		\label{fig:cost_functions}
	\end{figure*}

	Our sensitivity results are further supported by visualizing the \rss for both models {(FIG. \ref{fig:cost_functions})}. {The} \rss for scenario T3E is shown in FIG. \ref{fig:cost_functions}{(a)}, {and a greater variation in the \rss is visible in $\alpha$ relative to $\ve$}. Parameter values in the red shaded region fail to satisfy the steady-state condition, i.e., the model solution at \SI{20}{h} does not closely match the model solution at \SI{24}{h} (recalling that $\alpha$ controls the magnitude of the spatially-dependent velocity, if $\alpha$ is too small, then the retrograde velocity is too slow, and the solution takes longer than \SI{20}{h} to reach steady-state). However, for a complete visualization of the surface, we also plot the \rss without the steady-state condition underneath the red shaded area. This region, while not relevant to our analysis, reveals that our \rss is qualitatively smooth and exhibits no unusual nonlinearities. Parameter values in the white region result in an undefined spatial velocity profile $u(r)$. This issue arises because the spatial velocity profile \eqref{eq:u} is defined in terms of the reciprocal of the difference between the initial and final immobile vimentin, $\hat I$. If $\ve$ is sufficiently large, then there exists a zero crossing in $\hat I$, resulting in a singularity. Thus, numerically-computed solutions do not converge. {In contrast, the \rss of the best scenario T3F is always well-defined (FIG. \ref{fig:cost_functions}(c)). {We see a greater variation in the \rss surface in $\bar u$ relative to $\ve$, consistent with the sensitivity results}.}
	
	{Next, we compute the confidence interval for the estimates of $\alpha$ and $\ve$ for T3E and $\bar u$ and $\ve$ for T3F (for details, see Appendix \ref{a:confidence_interval}). The white curve in FIG. \ref{fig:cost_functions}(b) (resp. FIG \ref{fig:cost_functions}(c)) delimits the confidence interval for parameter values with a significance level of $p=0.05$ for T3E (resp. T3F). Solutions and the resulting approximates of the spatially-dependent velocity and net attachment rate obtained with parameters values from confidence intervals are shown in FIG. \ref{fig:best2sol}.}
	
	{Note that for the best scenario T3F, the confidence interval for the initial proportion of immobile vimentin $\ve$ is found to be \SI[separate-uncertainty = true]{55(6)}{\%}, which endorses the existence and requirement of a non-negligible proportion of vimentin that stays immobile in cells (FIG. \ref{fig:cost_functions}(c)).} {Furthermore, the confidence interval for $\bar{u}$ for T3F is found to be \SI[separate-uncertainty = true]{0.11(1)}{\um/min} (FIG. \ref{fig:cost_functions}(c)), which is in the range of the peak magnitude of retrograde velocity $u(r)$ in T3E that is on the order of \SI{0.1}{\um/\min} (FIG. \ref{fig:best2sol}(b)).} In similar studies of actin-dependent intermediate filament transport, magnitudes of retrograde velocity vary from \SI{0.16}{\um/\min} \cite{leduc2017regulation} to \SI{0.5}{\um/min} \cite{lehmann2021growth}. Our velocity estimates are consistent with \cite{leduc2017regulation}, \yp{but} our estimate may differ from \cite{lehmann2021growth}; however, this study was carried out in epithelial cells expressing keratin (another type of intermediate filament protein). Both differences in cell type and intermediate filament protein may explain the velocity difference. Moreover, we model data for \textit{net} retrograde flow of all intermediate filament material including {particles, single filaments or integrated in networks \ya{with slow or fast speeds}}. In contrast, the authors of \cite{lehmann2021growth} only follow individual particles that may not be homogeneous in velocity -- some particles exhibit \yp{movement in the} anterograde \yp{direction} and others \yp{in the} retrograde \yp{direction} -- and measure the average speed of transport independent of direction. Thus, their velocity estimates are greater than ours but still in the same order of magnitude.
	



\paragraph{Estimation of the velocity profile for actin-driven transport: }

While the agreement between T3E and the data is not as strong as scenario T3F, scenario T3E allows a spatially-dependent approximation of the velocity \eqref{eq:u} of retrograde actin-driven transport, by using vimentin material as an observable {(FIG. \ref{fig:best2sol}(a))}. As discussed above, the trapping rate $\alpha$ is the most influential parameter of the retrograde flow speed and controls the time to stabilization \yp{for intermediate filament material}.

\paragraph{Estimation of the net trapping rate profile: } 
Indeed, {scenario} T3F reproduces the data strikingly well at intermediate times. Another important contribution is that {scenario} T3F allows a spatially-dependent approximation of the net trapping rate \eqref{eq:alpha}, in which the retrograde velocity $\bar{u}$ is the most influential parameter, by using vimentin material as an observable {(FIG. \ref{fig:best2sol}(d))}.

\section{Discussion and Conclusion}

\yp{Here, we have combined experimental approach and mathematical \yp{modeling} to explore the mechanisms involved in the organization of the cytoplasmic \yp{intermediate filament} network \ya{in the context of no microtubule-driven transport}. By comparing \yp{10} different models using model identification, AIC model selection, and sensitivity analysis, we can identify two major biologically relevant models and extract valuable information including key parameters and the spatial distribution of biological  \yp{mechanisms}.}

\yc{Within the set of models considered,} the identifiable scenarios ( ``good models'' having single values \yc{or finite confidence intervals} for parameters) \yp{show that we must consider a pool of immobile vimentin filaments and that} immobile vimentin can not transition back to mobile vimentin (i.e., $\beta>0$ is not allowed). \yp{Our analysis also allows us to exclude scenarios}. If $\beta>0$, then all mobile vimentin eventually move out of the domain unless the retrograde velocity is zero, which contradicts the biological assumption of nonzero retrograde velocity. Similarly, if there is no transition from mobile to immobile, the only way to reach a non-zero steady-state is to either let the velocity be zero, or force all material to be immobile. Identifiable scenarios consistently show that $\beta=0$, {further suggesting that some mechanism for sequestering mobile vimentin is necessary.} 

\yp{We use model selection to quantify how well each scenario and its underlying assumptions and hypotheses reproduces the data.} \yp{Hence, we use the AIC model selection method to rank the scenarios}. Our model selection results \yp{bolster our confidence in our conclusions from model identifiability.} \yp{Indeed, under AIC, identifiable scenarios outperform non-identifiable {scenarios} when {reproducing} the data.} Furthermore, the \yp{top}-ranked {scenarios} in AIC are from {hypotheses D, E, and F}, where $\beta=0$, \yp{providing additional evidence that} irreversible trapping is an important feature of the {scenarios} we consider. In addition, we find that {scenario T3F}, irreversible trapping with spatially-dependent {trapping}, is unambiguously the best {scenario} out of the {scenarios} considered. The spatial profile of the trapping rate -- particularly where it increases towards the cell center -- is consistent with existing studies showing vimentin interaction with organelles, \yp{including the nucleus} \cite{gao2001novel,toivola2005cellular,chang2009dynamic,nekrasova2011vimentin,cremer2022vimentin}. Furthermore, recalling that $\alpha(r)$ in this case is a \textit{net} trapping rate, the negative values of $\alpha(r)$ towards the cell edge suggests that \yp{vimentin trapping is reduced at the cell periphery, possibly indicating a lower number of attachment sites \yc{for trapping}} \yp{(Fig. \ref{fig:best2sol}(d))}. 

{The second best scenario, T3E, which uses a spatially-dependent retrograde velocity, is also consistent with biological observations. In particular, the spatially-dependent velocity increases {from} the cell edge and decreases towards the cell center. This \yp{profile} suggests a few possible mechanisms} \yp{(Fig. \ref{fig:best2sol}(b))}. One, \yp{actin dynamics is locally regulated and varies from the cell periphery to the cell center, or} two, lower velocities correspond to regions with a high degree of crowding \yp{and increased drag}.

Our sensitivity results applied to both scenarios T3E and T3F show that they are sensitive to the transition rate from mobile to immobile, $\alpha$, and the constant retrograde velocity $\bar u$, respectively. Neither model is sensitive to the initial proportion of immobile vimentin, $\ve$. The latter observation tells us that the model can reproduce the data even if the initial data contains a small amount of immobile vimentin.

We highlight several important implications of our work. First, \yp{although it is tempting to explain experimental data (ours herein and \cite{hollenbeck1989intermediate}) by speculating that transport is solely sufficient to describe the observation, we conclude here that pure transport alone} (hypothesis \ref{hypothesis:transport}) insufficiently reproduces our data; we consistently find that hypothesis \ref{hypothesis:transport} \yp{can only} reproduce the data \yp{if a constant pool of  immobile vimentin is considered, which doesn't change in time}.

Second, the top two scenarios, T3E and T3F, both exhibit spatial dependence but in what appears to be a fundamentally related fashion. For example, the second best scenario, T3E, allows us to compute a spatially-dependent velocity $u(r)$, where its magnitude is directly proportional to the trapping rate $\alpha$. Interestingly, the best scenario, T3F, allows us to compute a spatially-dependent net trapping rate $\alpha(r)$, where its magnitude is directly proportional to the constant retrograde velocity $\bar u$. Due to the nature of our data, we are unable to determine the spatial profiles $\alpha(r)$ and $u(r)$ simultaneously. 

\yc{The above implications are immediately helpful to biologists who seek to understand the dynamics and the regulation of the intermediate filament network in cells -- both of which are still poorly understood. In particular, this modeling work enables experimentalists to pinpoint the biological questions that need to be addressed next. Prior to this paper, the plausibility of various mechanisms behind vimentin retrograde flow, e.g., whether or not there is a trapping mechanism or a non-constant retrograde velocity, was not known. The results of the present study suggest that vimentin retrograde flow is not a trivial advection, but may include some spatial variation along with a trapping mechanism. Moreover, we have demonstrated that the trapping rate could be spatially-dependent in such a way that it is possible for both the velocity and trapping rate to be spatially-dependent to some degree. This latter observation will be the subject of future experiments. For instance, by examining the spatial velocity-dependence of actin, researchers can gain insights into the corresponding spatial dependence of the vimentin trapping rate. In turn, this understanding could reveal the local molecular mechanisms that govern the trapping rate. Without the findings of this paper, there would be no basis for considering the trapping rate in subsequent experiments.}


\section*{Acknowledgements}
\yp{YP was supported by a PIMS Postdoctoral Fellowship \ya{(CTRMS-342044-2014)}. SEM and CL were supported by the Pasteur Institute (Paris, France) and the National Center for Scientific Research (CNRS).  SEM was supported by the La Ligue contre le cancer (S-CR17017). CL was supported by a French National Research Agency grant (ANR 16-CE13-019). SP was supported by a Discovery Grant of the Natural Sciences and Engineering Research Council of Canada (RGPIN-2018-04967) and a Burroughs Wellcome Fund 2020 Collaborative Research Travel Grant.}

\appendix

\section{Additional Method Details}
\subsection{Numerical Scheme}\label{a:numerical_scheme}
We use the following upwinding scheme to numerically integrate model \eqref{eq:model}:

\small
\begin{align*}
	I_i^{n+1} &= F_i^{n} + \Delta t T(M_i^n,I_i^n),\\
	M_i^{n+1} &= M_i^n + \Delta t \left[\frac{r_{i+1} u_{i+1} M_{i+1}^n - r_{i} u_i M_i^n}{r_i\Delta r} - T(M_i^n,I_i^n)\right],
\end{align*}
\normalsize
where $i=1,\ldots,N-1$, $T(M,I)=\yp{\alpha_i} M - \beta I$. The terms $\alpha_i$ and $u_i$ are defined as either constant or spatially-dependent. If the velocity is {quantity}-dependent, then we replace $u_i$ with the quantity-dependent function \eqref{eq:u_conc}
\begin{equation*}\label{eq:a:u_conc}
	u(V_i^n) = u_\text{m}\left(1 - \frac{V_i^n}{V_\text{m}}\right),
\end{equation*}
were $V_i^n=I_i^n+M_i^n$, and $u_m$, $V_m$ are free parameters representing the maximum retrograde velocity and maximum vimentin quantity respectively. We impose a Dirichlet condition on the right boundary: $M_{N+1}^n = 0$. \ya{As for numerical parameters, we use the time step \texttt{dt}=0.01 and mesh size \texttt{N}=100}.

\subsection{Model Selection }\label{sec:model_selection}
To select the best candidate {scenario} we use the Akaike information criterion (AIC). AIC selects for the {scenario} with the lowest $\rss$ while applying a penalty in the number of parameters \cite{akaike1974new}. Assuming independent and normally distributed additive measurement errors with the same variance, we may approximate the $\text{AIC}_i$ for model $i$ using the \rss:
\begin{equation*}
	\text{AIC}_i = N \ln\left(\frac{\rss_i{(\p_i^*)}}{N}\right) + 2K_i,
\end{equation*}
where $N$ is the number of observations and $K{_i}$ is the number of parameters (all parameters of {scenario} $i\in\mathcal{V}_s$ including one additional parameter from a bias correction term). We use 143 spatial points of the experimental data for each of the 7 time observations for a total of $N=1001$ observations.

The Akaike weights $w_i$ are then computed:
\begin{equation*}
	w_i = \frac{\exp(-\Delta_i/2)}{\sum_{j\in \mathcal{V}_s}\exp(-\Delta_j/2)},
\end{equation*}
where $\Delta_i = \text{AIC}_i - \min_{n\in\mathcal{V}_s} \left(\text{AIC}_n\right)$ is the difference between the AIC of {scenario} $i$ and the AIC of the {scenario} with the lowest AIC. Roughly speaking, the greater the Akaike weight $ w_i$, the stronger the evidence that {scenario} $i$ is the best {scenario} in the set of proposed {scenarios} \cite{portet2020primer}. 

\subsection{Confidence Interval}\label{a:confidence_interval}
\yp{For a given optimized parameter set $\p^*$, we compute} confidence intervals using the log-likelihood ratio statistic \cite{raue2009structural}, which provides an alternative method to approximating confidence intervals using the \rss. \yp{We approximate the log-likelihood of parameters by $\ln(\rss(\bm{\theta_0})/N)$, assuming independent and normally distributed additive measurement errors with the same variance.} Note that $N=1001$ is the total number of observations (143 spatial data points for each of the 7 observation times). Letting $\bm{\theta}_0=(\ve_0,{\alpha_0})$ {for T3E ($\bm{\theta}_0=(\ve_0,\yp{\bar u_{0}})$ for T3F)} denote an arbitrary parameter set and $\chi^2_{\gamma d}=7.815$ (where $\gamma = 5\%$ and $d=3$) the set defined by
\begin{equation}\label{eq:conf}
	\left\{\bm{\theta}_0: \ln\left[\frac{\rss(\bm{\theta}_0)}{N}\right] - \ln\left[\frac{\rss(\p^*)}{N}\right] \leq \frac{\chi^2_{\gamma d}}{N}\right\}
\end{equation}
defines the region of parameter space with a significance level at or below $p=0.05$. \yp{The boundary of the set \eqref{eq:conf} corresponds to the confidence interval.}

\begin{figure*}[htbp]
	\centering
	\includegraphics[width=.9\textwidth]{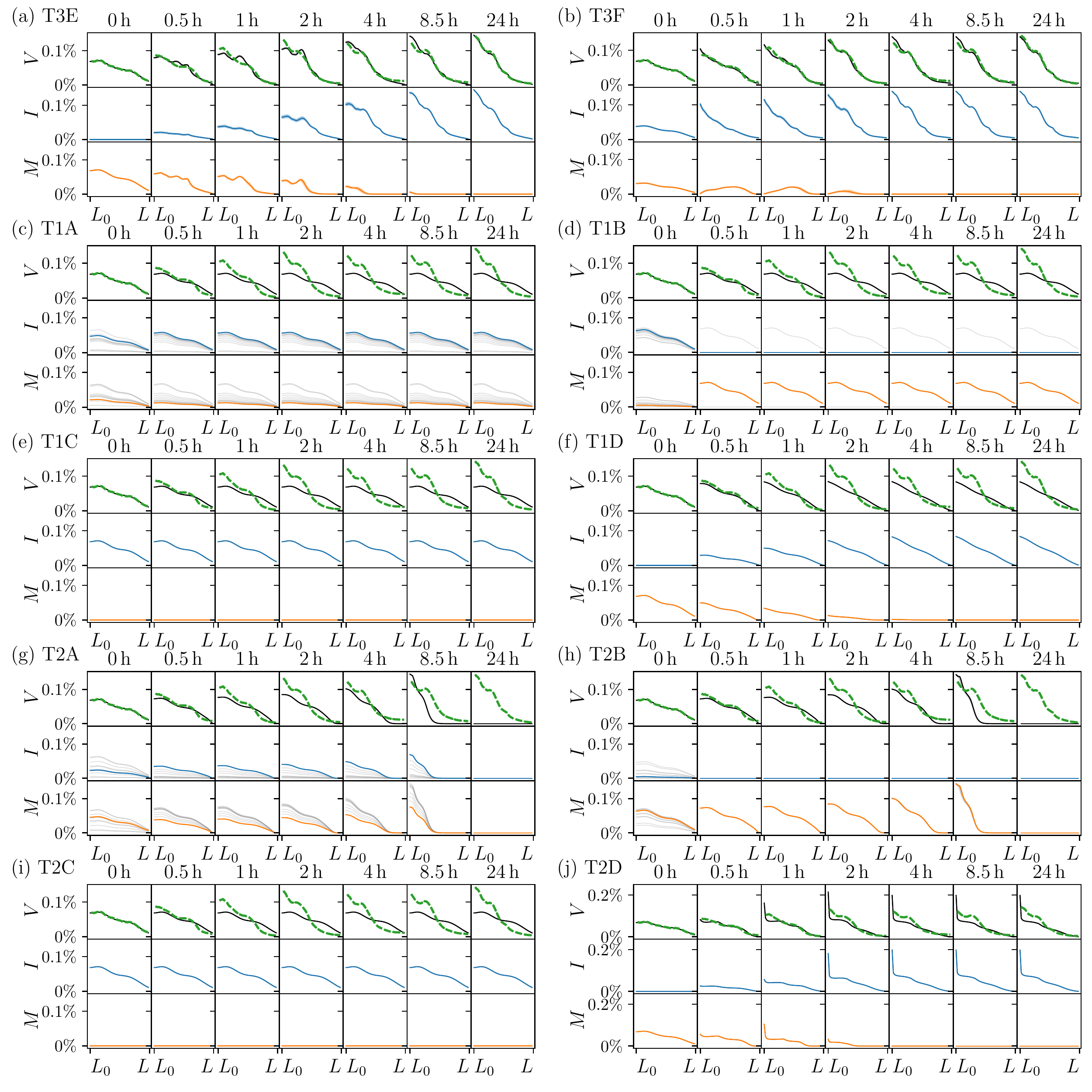}
	\caption{Solutions of all scenarios. The solution using the parameter set with the lowest \rss out of the 100 optimizations is plotted in black, blue, and orange for $V$, $I$, and $M$, respectively. Gray curves are solutions using parameters with the 9 next lowest \rss values. (a): Scenario {T3E}, {irreversible trapping} with spatially-dependent velocity. Optimized parameters: $\ve^*=0$, $\alpha^*=0.011$. (b): Scenario {T3F}, spatially-dependent {net} trapping rate {with constant velocity}. $\ve^*=0.55$, $\bar u^*=0.11$. (c): Scenario T1A{$^\dagger$}, trap and release {with constant velocity}. {$\ve^{*}=0.690$, $\alpha^{*}=19.205$, $\beta^{*}=4.180$, $\bar u^{*}=\num{6.175e-7}$}. (d): Scenario {T1B}{$^\dagger$}, only release with no trapping {with constant velocity}. {$\ve^{*}=0.925$, $\beta^{*}=0.562$, $\bar u^{*}=\num{1.098e-7}$}. (e): Scenario T1C$^\dagger$, pure transport {with constant velocity}, ($\alpha=\beta=0$). {$\ve^{*}=1$, $\bar u^{*}=\num{6.189e-4}$.} (f): Scenario {T1D}, {irreversible} trapping {with constant velocity} ($\beta=0$). $\ve^{*}=0$, $\alpha^{*}=1.657\times10^{-2}$, $\bar u^{*}=5.400\times10^{-2}$. (g): Scenario {T2A}{$^\dagger$}, trap and release {with vimentin-dependent velocity} {$\ve^{*}=0.335$, $V_\text{m}^{*}=\num{1.167e-3}$, $u_\text{m}^{*}=\num{5.277e-2}$, $\alpha^{*}=10.527$, $\beta^{*}=11.537$} (h): Scenario {T2B}{$^\dagger$}, only release with no trapping {with vimentin-dependent velocity} ($\alpha=0$).  $\ve^{*}=\num{6.981e-2}$, $V_\text{m}^{*}=\num{2.131e-3}$, $u_\text{m}^{*}=\num{2.765e-2}$, $\beta^{*}=6.549$. (i): Scenario {T2C}{$^\dagger$}, pure transport {with vimentin-dependent velocity} ($\alpha=\beta=0$). {$\ve^{*}=1.0$, $V_\text{m}^{*}=0.549$, $u_\text{m}^{*}=\num{2.953e-3}$}. (j): Scenario {T2D}, {irreversible trapping with vimentin-dependent velocity} ($\beta=0$). {$\ve^*=0$, $V_\text{m}^*=\num{6.599e-6}$, $u_\text{m}^*=0.111$, $\alpha^*=1.477\times10^{-2}$}.  {$^\dagger$}A dagger indicates a non-identifiable scenario.}
	\label{fig:sols}
\end{figure*}

\section{Derivation of Spatially-Dependent Terms}\label{a:spatially-dependent}

We consider the special case where there is no explicit detachment ($\beta=0$) and allow $\alpha$ or $u(\cdot)$ (but not both) to be nonconstant in space (resp. hypotheses E and F):
\begin{align*}
	\frac{\partial I}{\partial t}= &   \alpha(r) {M},\\
	\frac{\partial M}{\partial t}= &  \frac{1}{r}\frac{\partial}{\partial r} (ru(r) M) -\alpha(r) {M},\\
	I(r,0) = \ve\tilde V_\text{0}(r), &\quad M(r,0) = (1-\ve)\tilde V_\text{0}(r).
\end{align*}

\noindent Plugging the first equation into the second yields
\begin{equation*}
	\frac{1}{\alpha(r)}\frac{\pa^2 I}{\pa t^2} = \frac{1}{r}\frac{\pa}{\pa r}\left( \frac{r u(r)}{\alpha(r)} \frac{d{I}}{dt}\right) - \frac{\pa I}{\pa t}.
\end{equation*}

\noindent Then integrating with respect to time results in a first-order differential equation,
\begin{equation}\label{eq:df_c}
	{\frac{1}{\alpha(r)}} \frac{\pa I}{\pa t}(r,t) = \frac{1}{r}\frac{\pa}{\pa r}\left( \frac{r u(r)}{\alpha(r)} I(r,t)\right) - I(r,t) + c(r),
\end{equation}

\noindent where $c(r)$ is a spatially-dependent constant of integration. To solve for $c(r)$, let $t=0$ in \eqref{eq:df_c}:
\begin{align*}
	\frac{1}{{\alpha(r)}}\frac{\pa I}{\pa t}(r,0)& \equiv  M(r,0)\\
	&= \frac{1}{r} \frac{\pa}{\pa r} \left(\frac{ru(r)}{\alpha(r) } I(r,0)\right) - I(r,0) + c(r),
\end{align*}

\noindent then solve for $c(r)$ directly:
\begin{equation*}
	c(r) = M_0(r) + I_0(r) - \frac{1}{ r} \frac{\pa}{\pa r} \left(\frac{ru(r)}{\alpha(r)} I_0(r)\right),
\end{equation*}

\noindent where $I_0(r) := I(r,0):= \yp{\ve \tilde V_{0}(r)}$ and $M_0(r) := M(r,0) := \yp{(1-\ve)\tilde V_0(r)}$. We plug $c(r)$ back into \eqref{eq:df_c} and take $t \rightarrow \infty$:
\begin{equation}\label{eq:ss}
	0 = - \frac{1}{r} \frac{\pa}{\pa r} \left(\frac{r u(r)}{\alpha(r)} {\hat I(r)}\right) + \hat I(r) + M_0(r),
\end{equation}

\noindent where $\hat I(r) := I_0(r) - I^*(r)$ with $I^*(r) := \tilde V_{24}(r)$ (as we assume that at 24h the vimentin profile has reached its steady state and all vimentin is immobile). We use the ODE \eqref{eq:ss} to derive the spatially-dependent terms $u(r)$ or $\alpha(r)$.

\subsection{Spatially-Dependent Velocity}\label{a:u}

Consider \eqref{eq:ss} with $\alpha > 0$ constant and $u(r)$ spatially-dependent. Then the ODE for $u(r)$ is given by
\begin{equation}\label{eq:u_ode}
	{\frac{\pa }{\pa r}\left( r u(r)\hat I(r)\right) = \alpha r \left[\hat I(r) + M_0(r)\right]},
\end{equation}

\noindent which has the solution,
\begin{equation}\label{eq:a_u}
	u(r) = \frac{\alpha}{r\hat I(r)} \int_{L_0}^r s\left(\hat  I(s)  + M_0(s)\right) \,\mathrm{d}s + \frac{L_0 u(L_0) \hat I(L_0) }{r\hat I(r)}.
\end{equation}
\yp{We} assume $u(L_0) = 0$ to enforce conservation of mass. Note that the velocity equation \eqref{eq:a_u} depends explicitly on $\alpha$, the transition rate of mobile to immobile vimentin, and implicitly on $\ve$, the proportion of immobile vimentin (because the initial conditions $M_0$ and $I_0$ depend on $\ve$). Therefore, using the spatially-dependent velocity with conservation of mass \yp{\eqref{eq:a_u}} results in a two-parameter model.

\subsection{Spatially-Dependent Net Trapping Rate}\label{a:alpha}

Again consider {\eqref{eq:ss}}, but now assume $u(r) = \bar u$ and that $\alpha(r)$ is spatially-dependent. Then the ODE for $\alpha(r)$ given by,
\begin{equation*}
	{\frac{\pa}{\pa r}\left( \frac{r}{\alpha(r)}\hat I(r) \right) = \frac{r}{\bar u}\left[\hat I(r) + M_0(r)\right]},
\end{equation*}

\noindent which has the solution,
\begin{equation*}
	\alpha(r) = \frac{r\bar{u}\hat I(r)}{\int_{L_0}^r s(\hat I(s) + M_0(s)) \,\mathrm{d}s - L_0 \bar u \hat I(L_0)/\alpha(L_0)}.
\end{equation*}

\noindent To ensure conservation of mass, we assume that $\alpha(L_0)$ is large, thus we may approximate the \yp{spatially-dependent net trapping rate} with the simpler equation,
\begin{equation*}
	\alpha(r) = \frac{r\bar{u}\hat I(r)}{\int_{L_0}^r s(\hat I(s) + M_0(s)) \,\mathrm{d}s}.
\end{equation*}

\section{Best Fit Solutions}\label{a:solutions}

For each {scenario} $i \in \mathcal{V}_s$, its ``optimized parameters'' simply refer to the parameters that result in solutions that optimally fit the data according to the  \yp{optimization error} \eqref{eq:opt}. The optimal parameters are found using the inherently stochastic differential evolution method, which we run 100 times for each {scenario} to ensure that we find a global minimum in parameter space. We then plot the solutions corresponding to the {10 optimized parameter sets with the lowest \yp{optimization error} values in FIG. \ref{fig:sols}}. The best of the 10 optimized parameter sets (according to the \yp{optimization error}) are plotted in black (total vimentin $V_i=I_i+M_i$), blue (immobile vimentin $I_i$), and orange (mobile vimentin $M_i$). Solutions corresponding to the remaining 9 runs are shown in gray.


\end{document}